\documentclass[journal]{IEEEtran}
\usepackage{graphicx, subfigure}
\usepackage{commath}
\usepackage{mathtools}
\usepackage{amsmath}
\usepackage{cite}
\usepackage{color}
\usepackage{array}
\usepackage{booktabs}
\usepackage{tabularx}

\ifCLASSINFOpdf
\else
\fi

\begin{document}
%
\title{Analysis of MEMS electrostatic energy harvesters electrically configured as voltage multipliers}
%
%
%

\author{Binh~Duc~Truong,~Cuong~Phu~Le
        and~Einar~Halvorsen,~\IEEEmembership{Member,~IEEE}
\thanks{B. D. Truong, C. P. Le and E. Halvorsen are with the Department
of Microsystems, University College of Southeast Norway, e-mail: Binh.Truong@usn.no.}
\thanks{Manuscript received Moth Day, Yeah; revised Moth Day, Year.}
}

\maketitle

\begin{abstract}
This paper presents the analysis of an efficient alternative interface circuit for MEMS electrostatic energy harvesters. It is entirely composed by diodes and capacitors. Based on modeling and simulation, the anti-phase gap-closing structure is investigated. We find that when configured as a voltage multiplier, it can operate at very low acceleration amplitudes. In addition, the allowed maximum voltage between electrodes is barely limited by the pull-in effect. The parasitic capacitance of the harvester and non-ideal characteristics of electronic components are taken into account. A lumped-model of the harvesting system has been implemented in a circuit simulator. Simulation results show that an output voltage of 22 V is obtained with 0.15 g input acceleration.  The minimum necessary ratio between the maximum and minimum capacitances of the generators which allows operation of the circuit, can be lower than 2. This overcomes a crucial obstacle in low-power energy harvesting devices. A comparison between the voltage multiplier against other current topologies is highlighted. 
An advantage of the former over the latter is to generate much higher saturation voltage, while the minimum required initial bias and the minimum capacitance ratio in both cases are in the similar levels.
\end{abstract}


%
\IEEEpeerreviewmaketitle

\section{Introduction}

\IEEEPARstart{V}{ibration-based}
energy harvesting can provide reliable alternative power sources for emerging miniaturized-electrical devices such as implant medical/wireless sensors \cite{Roundy2004, Mitcheson2010}. As a key element connecting the harvester and storage component or electrical load, power electronic interface circuitry has gained significant interest in recent years. Abundant research on circuit topology with functions of voltage regulation, optimal power extraction and damping control for energy harvesting systems has been conducted and presented \cite{Ottman2003, Cao2007, Mitcheson2007, Zuo2009, Dhulst2010}. In this paper, we focus on the power electronics for micro capacitive energy harvester.

Several topologies of power electronic interface circuit for MEMS electrostatic energy harvesters are reported. For instance, Yen \textit{et al.} proposed a simplified circuit in which a charge pump was used with a fly-back circuit to store and transfer extracted energy \cite{Yen2006}. The major drawback of this architecture is that the storage capacitor acts as parasitic capacitance to the energy converter, reducing the net power harvested. The power consumption of the control unit of the switch is also a challenging concern as it must be supported by the generator, even in ASIC implementations \cite{Dudka2013, Phan2015}. In addition, the use of bulky inductor is not applicable to miniaturization of the harvesting system.

Besides the switched-inductor topologies, diode-capacitor voltage multipliers are widely used for converting an AC input voltage into a DC output voltage.
Based on Bennet's doubler of electricity, de Queiroz \textit{et al.} proposed an inductor-less and switch-less doubler circuit \cite{deQueiroz2011, deQueiroz2013}. The major advantage of this topology is a simple and robust implementation. However, the circuit requires a ratio $C_\mathrm{max}/C_\mathrm{min}$ between maximum and minimum capacitances of the variable capacitors to be greater than 2. This is a critical challenge in practice since MEMS harvesters tend to exhibit small variable capacitances, limited by standard fabrication processes and device size.

An advanced topology for vibration electrostatic energy harvesters was recently proposed by Lefeuvre \textit{et al.} \cite{Lefeuvre2014, Wei2015}. An argument based on theoretical analysis using a rectangular Q-V cycle suggests that this voltage multiplier can operate with a minimum necessary capacitance ratio lower than 2. However, one may have concerns regarding the accuracy of this assumption for the complex dynamic behavior of the circuit. Also, both simulations and experiments were for a macro-scale rotating variable-capacitor with $C_\mathrm{min}=45$ pF and $C_\mathrm{max}=155$ pF which clearly has $C_\mathrm{max}/C_\mathrm{min}$ much larger than 2.

This paper presents a new adaption of a voltage multiplier circuit to MEMS energy harvesting. This capacitor-diode network is based on the Greinacher's voltage doubler which is the basic stage of a multiplier circuit first proposed by Greinacher in 1920 \cite{Hauschild2014}.
Since the micro energy harvesting system typically provides low output voltage that is not acceptable to switching converters, the multiplier topology is also a potential solution to initially boost such voltage to start-up the active elements of converter circuit \cite{Dayal2011, deQueiroz2011b, Wei2016}.
A complete lumped model including both mechanical and electrical domains is investigated. Parasitic capacitance of the generators and diode losses are taken into account. System dynamics are analyzed using a SPICE simulator.
The performance of the introduced multiplier configuration is then compared to several circuits of the same family \cite{Lefeuvre2014, Karami2017}.

\section{Energy harvester model}



\begin{figure}[!htbp]
	\centering
	\includegraphics[width=0.45\textwidth]{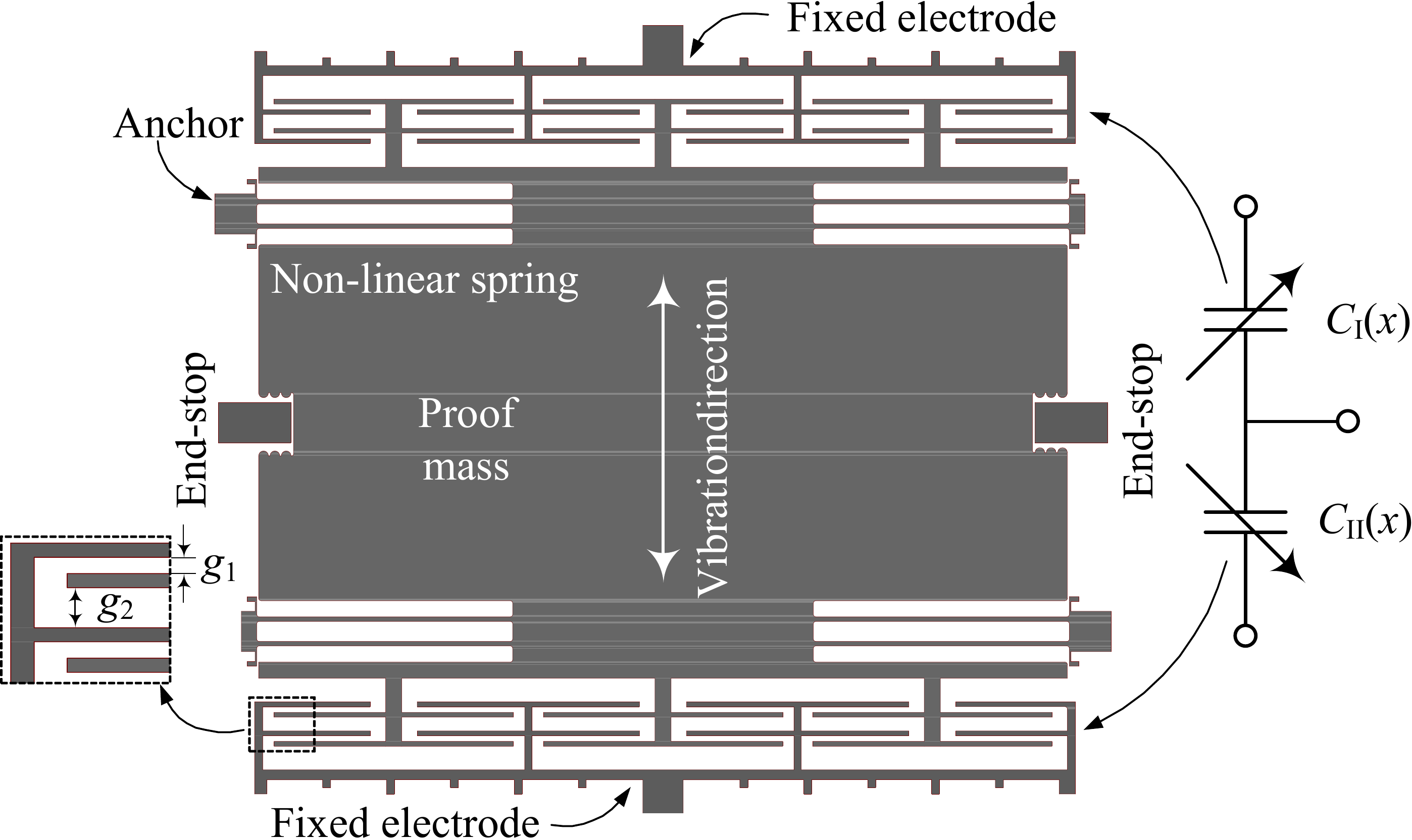}
	\caption{Anti-phase gap-closing vibration energy harvester.}
	\label{fig_EH}
\end{figure}
Figure \ref{fig_EH} shows the 2D geometry of the electrostatic energy harvester, which is used as a vehicle to explore the operation of a voltage multiplier configuration. The device includes two fixed electrodes attached to the frame and two variable electrodes on the proof mass suspended by four non-linear springs. The maximum displacement $X_\mathrm{max}$ is limited by mechanical end-stops. Though an in-plane anti-phase gap-closing asymmetric transducer is early used in sensing applications \cite{Sun2005, KKim2008}, it is the first time introduced as an energy generator through this paper. An advantage of such structure is the capability to obtain large range of capacitance variation with small displacement, which allows the device to operate efficiently at relatively low input excitations. Meanwhile, the pull-in voltage is kept to an remarkably high value due to the anti-phase behavior of the two transducers. These characteristics make the structure interesting to investigate.

The capacitances $C_\mathrm{I}$ and $C_\mathrm{II}$ as functions of the proof mass displacement are
\begin{eqnarray} \label{Eq_Cap}
C_\mathrm{I/II} \big(x\big) = C_\mathrm{p} + C_\mathrm{1/2}  \big(x\big), \\
C_1  \big(x\big) = C_0   \frac{1}{\big( 1 +  \frac{x}{g_1} \big)  \big( 1 -  \frac{x}{g_2}  \big) } , \\
C_2  \big(x\big) = C_0   \frac{1}{\big( 1 -  \frac{x}{g_1} \big)  \big( 1 + \frac{x}{g_2}  \big) }
\end{eqnarray}
where $C_0$ is the nominal capacitance of transducers, $C_\mathrm{p}$ presents for the parasitic capacitance. The gaps between movable electrode and fixed electrodes $g_1$ and $g_2$ must be satisfied the condition $X_\mathrm{max} < g_1 < g_1 + 2 X_\mathrm{max} < g_2$ so that the anti-phase characteristics is fulfilled $\frac{\partial C_\mathrm{I}}{\partial x} \frac{\partial C_\mathrm{II}}{\partial x} < 0$.

The nonlinear squeeze-film damping is proved to be essential to capture the device behavior \cite{Kaur2015, Truong2015}. Meanwhile, the hardening restoring force could be useful to increase the device stability during operation.
Along with the power conversion circuitry, the effect of the mechanical end-stops on the performance of a harvesting system is significant and cannot be ignored \cite{Blystad2010}.
Therefore, all of them needs to be taken into account in device modeling.
The dynamic of the spring-mass-damper system is then given by
\begin{eqnarray} \label{Eq_Newton}
m  \ddot{x} + B \big(x\big)  \dot{x} + F_\mathrm{m} \big(x\big) + F_\mathrm{e} \big(x\big) + F_\mathrm{s} \big(x\big) = F
\end{eqnarray}
where $m$ is the inertial mass and $F=mA \cos \big(\omega t\big)$ is the external force.
Including the squeeze-film damping of the gap-closing capacitor structure \cite{Bao2007} and an additional linear damping, the total damping force of the harvesters is $B \big(x\big)  \dot{x}$, where
\begin{eqnarray} \label{Eq_Damping}
	\begin{aligned}
		B(x) = b_\mathrm{l} +  \frac{b_\mathrm{n}}{ \big( 1 +  \frac{x}{g_1} \big) ^3} + \frac{b_\mathrm{n}}{ \big( 1 -  \frac{x}{g_2} \big) ^3} \\
		+ \frac{b_\mathrm{n}}{ \big( 1 -  \frac{x}{g_1} \big) ^3} + \frac{b_\mathrm{n}}{ \big( 1 +  \frac{x}{g_2} \big) ^3}.
	\end{aligned}
\end{eqnarray}
$F_\mathrm{m}$ is the mechanical restoring force, which we consider on the Duffing-spring polynomial form 
\begin{eqnarray} \label{Eq_Fm}
	F_\mathrm{m} \big( x \big) = k_\mathrm{1} x + k_\mathrm{3} x^3.
\end{eqnarray}
The electrostatic force is represented by
\begin{eqnarray} \label{Eq_Fe}
	F_\mathrm{e} \big( x \big) =  \frac{1}{2}  q_{1}^2  \frac{\mathrm{d}}{\mathrm{d} x}  
	\frac{1}{C_1 \big(x\big) + C_\mathrm{p}} 
	+ \frac{1}{2}  q_{2}^2  \frac{\mathrm{d}}{\mathrm{d} x}  
	\frac{1}{C_2 \big(x\big) + C_\mathrm{p}} 
\end{eqnarray}
where $q_1$ and $q_2$ are the charges on the two transducers. A stray capacitance $C_\mathrm{p}$ in parallel with the variable capacitance is also included in the model. 
$F_\mathrm{s}$ is the impact force between the movable mass and the rigid end-stops at the ultimate displacement $\abs x > X_\mathrm{max}$, which can be simplified as \cite{Le2012}
\begin{eqnarray} \label{Eq_Fs}
	F_\mathrm{s} \big( x \big) = k_\mathrm{s} \delta + b_\mathrm{s}  \dot{\delta}
\end{eqnarray}
where $k_\mathrm{s}$ is the impact stiffness, $b_\mathrm{s}$ is the damping coefficient accounting for the impact loss and $\delta = \abs x - X_\mathrm{max}$ is the deformation displacement between the proof mass and the end-stops during impact. 

\begin{figure}[!t]
	\centering
	\includegraphics[width=0.34\textwidth]{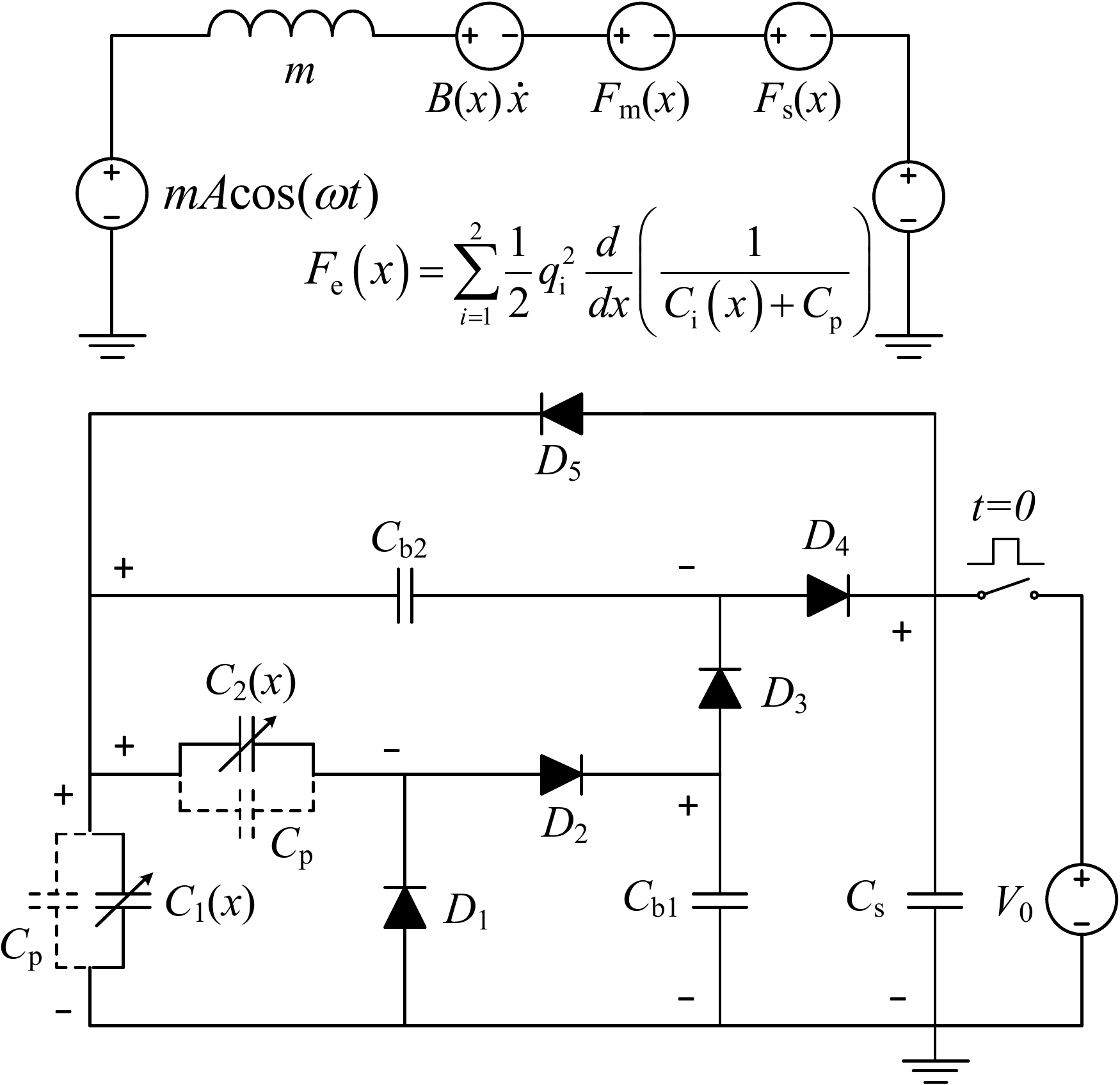}
	\caption{Lumped-model of electrostatic energy harvester configured as a voltage multiplier.}
	\label{fig_Model}
\end{figure}
Based on the above equations, the equivalent circuit for the anti-phase gap-closing electrostatic energy harvesters is shown in Figure \ref{fig_Model}. Along with that is the proposed interface circuit, which includes two, namely, biasing capacitor $C_\mathrm{b_1}$ and $C_\mathrm{b_2}$ and five low-leakage diodes $D_1,\ D_2, \, D_3, \, D_4$ and $D_5$ (abbreviation, $D_{1-5}$). The output of the interface circuit is connected to a storage capacitor $C_\mathrm{s}$ where the generated energy is stored. $C_\mathrm{s}$ is initially precharged to a voltage of $V_0$. All model parameters are listed in Table \ref{Tab_Params} and are chosen close to the parameters of the MEMS electrostatic energy harvesters in \cite{Nguyen2010JMM}.

\begin{table}
	\centering
	\caption{Model parameters}%
	\begin{tabular}{l l}
		\hline\hline
		\textbf{Parameters} & \textbf{Values} \\
		\hline
		Proof mass, $m$ & 30.4 mg \\
		Linear stiffness, $k_1$ & 675.1 N$\mathrm{m}^{-1}$ \\
		Hardening nonlinear stiffness, $k_3$ & 2.7e7 N$\mathrm{m}^{-3}$ \\
		Slide-film air damping, $b_\mathrm{l}$ & 2.5e-5 Ns$\mathrm{m}^{-1}$ \\
		Squeeze-film air damping, $b_\mathrm{n}$ & 2.7e-5 Ns$\mathrm{m}^{-1}$ \\
		Nominal capacitance, $C_0$ & 36.26 pF \\
		Parasitic capacitance, $C_\mathrm{p}$ & 8.5 pF \\
		Lower gap between movable and fixed electrodes, $g_1$ & 16 $\mu$m \\
		Upper gap between movable and fixed electrodes, $g_2$ & 40 $\mu$m \\
		Maximum displacement, $X_\mathrm{max}$ & 6.5 $\mu$m \\
		Impact stiffness, $k_\mathrm{s}$ & 3.36e6 N$\mathrm{m}^{-1}$ \\
		Impact damping, $b_\mathrm{s}$ & 0.44 Ns$\mathrm{m}^{-1}$ \\
		Bias capacitance, $C_\mathrm{b1/b2}$ & 0.5 nF \\
		Storage capacitance, $C_\mathrm{s}$ & 10 nF \\
		\hline\hline
	\end{tabular}
	\label{Tab_Params} 
\end{table}

\section{Operation and dynamics of the two-stage voltage multiplier} \label{Two-stage}

\begin{figure}[!t]
	\centering
	\includegraphics[width=0.385\textwidth]{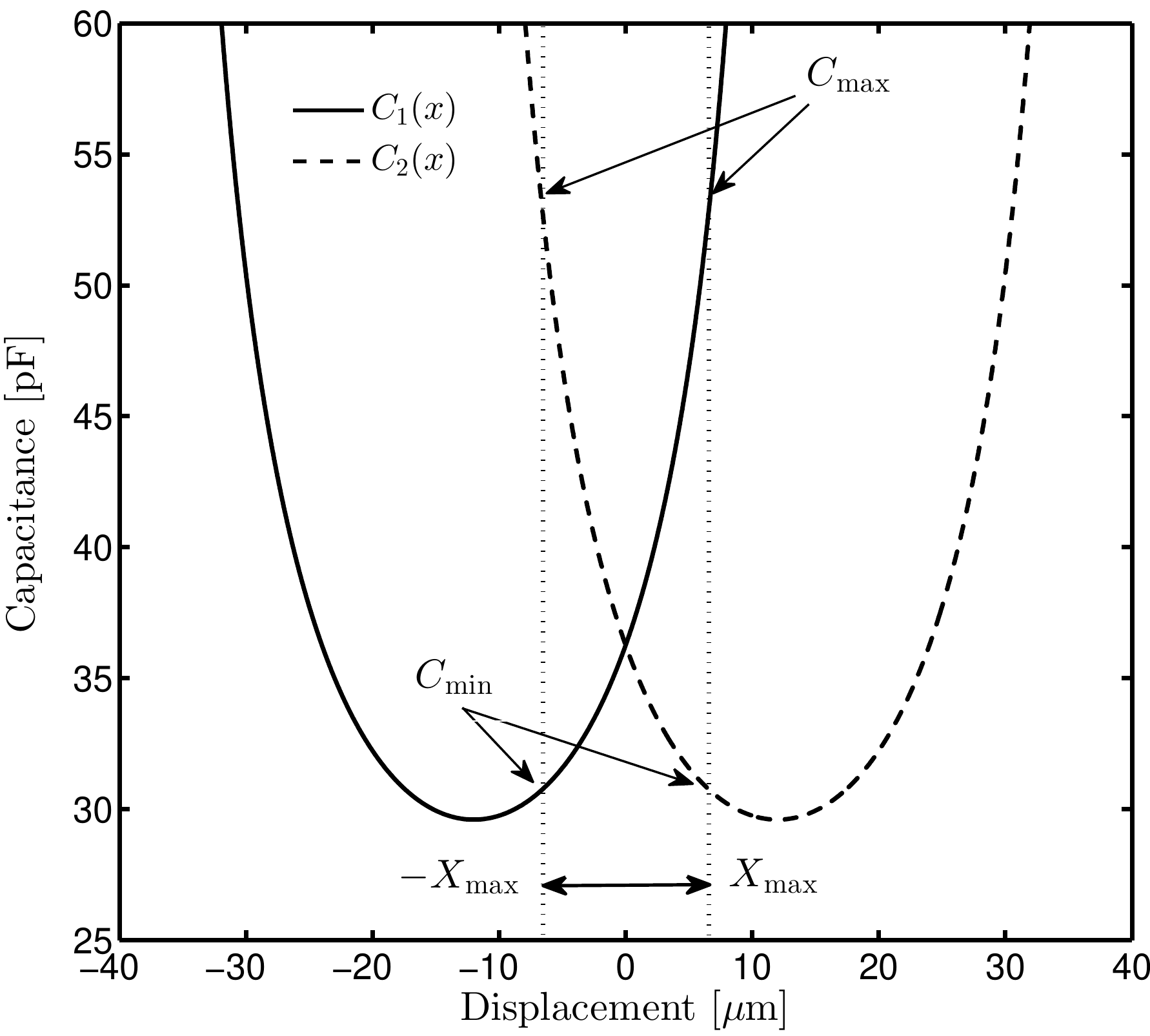}
	\caption{The capacitance variation in anti-phase gap-closing transducers.}
	\label{fig_Cap}
\end{figure}

In order to investigate the potential benefit of the circuit, the MEMS transducers are designed so that the capacitance variation ratio is
\begin{eqnarray}
\eta = \frac{C_\mathrm{max} + C_\mathrm{p}}{C_\mathrm{min} + C_\mathrm{p}} = 1.55.
\end{eqnarray}
It is also the minimum ratio that is required for the harvester to work as a voltage multiplier. Note that the critical ratio for operation of the Bennet's doubler is $\eta_\mathrm{cr}=2$, or even larger when the electrical losses are included.
The variable capacitors here vary between $C_\mathrm{min}=30.79$ pF and $C_\mathrm{max}=52.53$ pF. The capacitance variation of the two transducers are shown in Figure \ref{fig_Cap}.

A dynamic model of the low-leakage diode BAS716 \cite{BAS716diode} is used for simulations in LT-SPICE, taking into account non-ideal properties such as built-in junction voltage, leakage current and \textit{p-n} junction capacitance. The SPICE model is much more complicated and realistic than constant-voltage-drop model or piecewise linear model. The capacitance value of the storage capacitor is first arbitrarily chosen large enough $C_\mathrm{s} = 10$ nF to minimize the ripple voltage at the DC output, while the biasing capacitance is kept much smaller $C_\mathrm{b1} = C_\mathrm{b2} = C_\mathrm{b} = 0.5$ nF to minimize start-up time.

\begin{figure}[!htbp]
	\centering
	\includegraphics[width=0.4\textwidth]{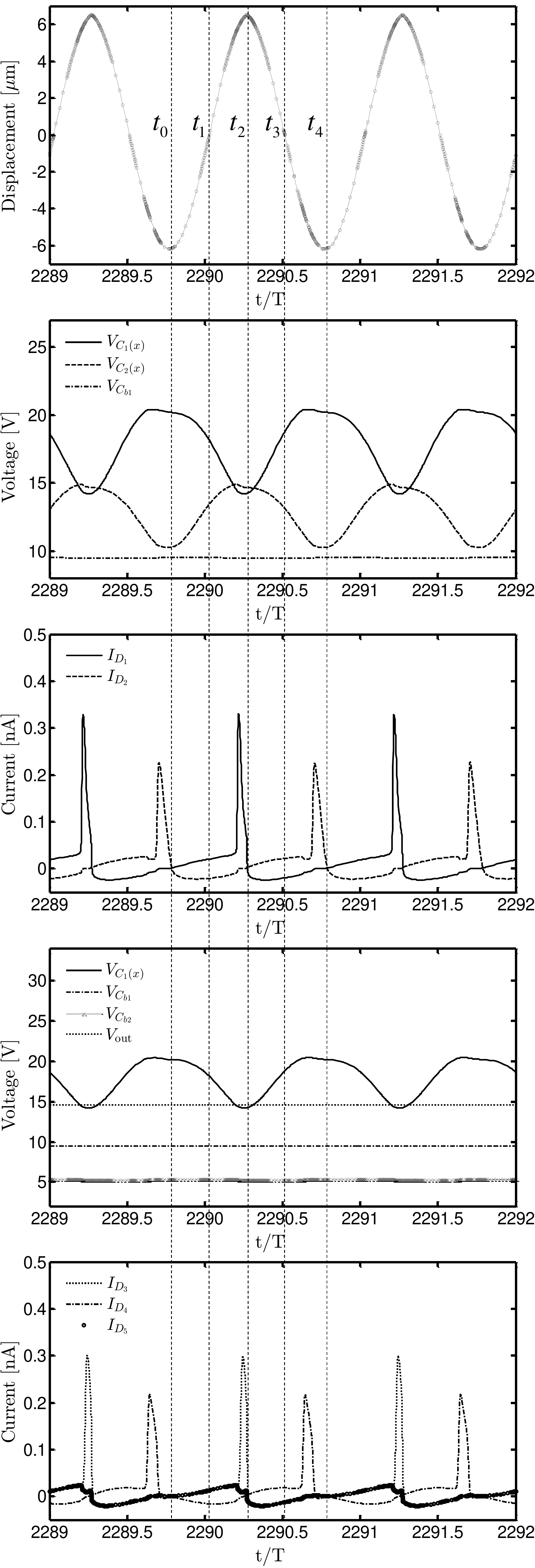}
	\caption{Simulated waveforms of the proof mass displacement, voltages across capacitors and currents through diodes. The input acceleration amplitude is $A=0.5$ g. The driving frequency is kept at the constant value $f_0 = \sqrt{ \frac{k_1}{m}} = 750$ Hz and the initial bias voltage is $V_0 = 12$ V.}
	\label{fig_Operation}
\end{figure}
Figure \ref{fig_Operation} shows waveforms of the voltages across capacitors and currents through diodes together with the proof mass displacement in the same time interval. The voltages across $C_\mathrm{b1/b2}$ and $C_\mathrm{s}$ increase slowly as functions of period of time, while that of variable capacitors have AC characteristics depending on the amplitude and frequency of the input acceleration. Use of the non-ideal diode model results in strongly complex behavior of the diode currents.
To analyze the complicated operation of the circuit, a sequence of four stages can be identified during a cycle.

In the first stage, from $t_0$ to $t_1$, $C_1(x)$ $\big(C_2(x)\big)$  start increasing (decreasing) from the minimum (maximum) value. Conditions $V_\mathrm{C_2} + V_\mathrm{C_\mathrm{b1}} > V_\mathrm{C_1} > V_\mathrm{C_2}$, $V_\mathrm{C_1} > V_\mathrm{C_\mathrm{b1}} + V_\mathrm{C_\mathrm{b2}}$ and $V_\mathrm{out} + V_\mathrm{C_\mathrm{b2}} > V_\mathrm{C_1} > V_\mathrm{out}$ are satisfied. All the diodes are blocked and the charges of $C_1$ and $C_2$ can be treated as constants.

The second stage is from $t_1$ to $t_2$. At the beginning, $V_\mathrm{C_1}$ continues decreasing while $V_\mathrm{C_2}$ is increasing. When $V_\mathrm{C_2} > V_\mathrm{C_1}$, $D_1$ is conducting and $C_1$ is charged by $C_2$. Also, $V_\mathrm{C_\mathrm{b1}} + V_\mathrm{C_\mathrm{b2}} > V_\mathrm{C_1}$ leads to conduction of the diode $D_3$ and the current flows from $C_\mathrm{b1}$, $C_\mathrm{b2}$ to $C_1$. This stage ends when $C_1$ reaches the minimum value $C_\mathrm{min}$ and $V_\mathrm{C_1}$ starts to rise.

In the next stage from $t_2$ to $t_3$, all diodes $D_{1-5}$ are reverse-biased, for the same reasons as above. However, $C_1$ now is decreasing from the maximum value $C_\mathrm{max}$ and the voltage across it increases again. The opposite happens to $C_2$.

In the final stage from $t_3$ to $t_4$, $V_\mathrm{C_1}$ first continues increasing while $V_\mathrm{C_2}$ decreases. This situation prevails until $V_\mathrm{C_1}$ becomes larger than $V_\mathrm{C_2} + V_\mathrm{C_\mathrm{b1}}$ and the diode $D_2$ begin to conduct transferring charge from  $C_1$ and $C_2$ to $C_\mathrm{b1}$. Then, the voltage across $C_1$ changes only very slightly. In addition, when $V_\mathrm{C_1} > V_\mathrm{out} + V_\mathrm{C_\mathrm{b2}}$, the current flows from $C_1$ and  $C_\mathrm{b2}$ to $C_\mathrm{s}$ while $D_4$ is conducting. When $C_1$ reaches $C_\mathrm{max}$, all conditions mentioned in the first stage occur and a new cycle starts. Note that the condition $V_\mathrm{out} > V_\mathrm{C_1}$ is rarely satisfied. The current that transfers the charge from $C_\mathrm{s}$ to $C_1$ through the diode $D_5$ is small and negligible. However, $D_5$ still plays a vital role for pre-charging $C_1$ at the very beginning of operation. In principle, $D_5$ can be disconnected through a switch in series after the first several cycles. This technique could result in higher saturation voltage \cite{TruongEHS2017}, but is not our objective in this paper. Therefore, we choose to keep $D_5$ as part of the voltage multiplier, so that the attractive simplicity of the circuit is not compromised.

\begin{figure}[!t]
	\centering
	\includegraphics[width=0.385\textwidth]{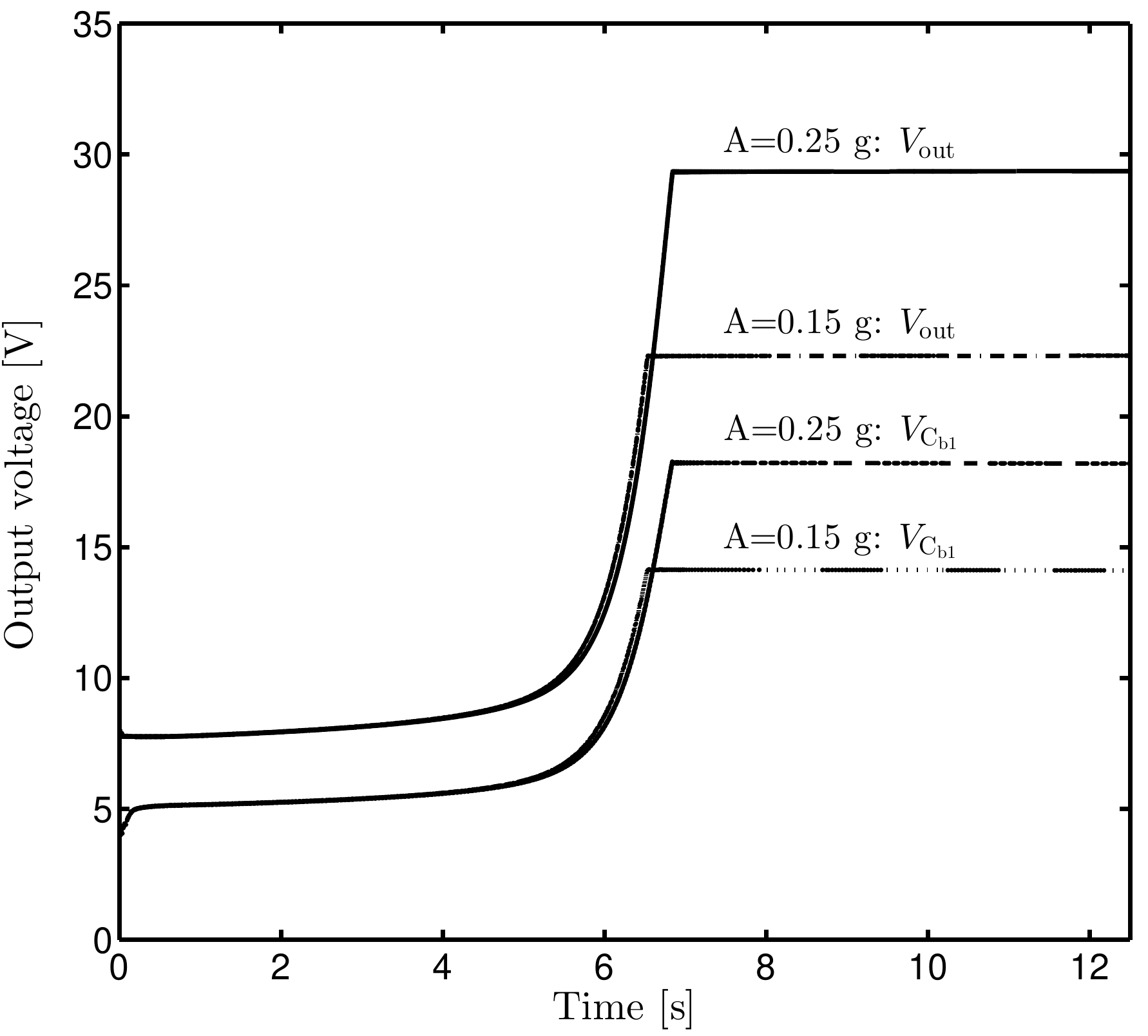}
	\caption{The time evolution of the voltage across $C_\mathrm{b1}$ and $C_\mathrm{s}$ denoted as $V_\mathrm{C_1}$ and $V_\mathrm{out}$ respectively, for different input acceleration $A=0.15$ g and $A=0.25$ g.}
	\label{fig_Vout}
\end{figure}
Figure \ref{fig_Vout} shows the time evolution of the output voltage with drive frequency $f=f_0 = \frac{1}{2 \pi} \sqrt{\frac{k_1}{m}} = 750$ Hz, acceleration amplitudes $A=0.15$ g and $A=0.25$ g and initial pre-charge voltage $V_0 = 8$ V. The energy stored in $C_\mathrm{s}$ initially increases. The steady state is obtained after several cycles of operation and the accumulated voltage $V_\mathrm{out}$ is saturated at a certain level due to the effect of the electromechanical coupling \cite{TruongSensors2017a}. Higher voltages result in stronger electrostatic forces causing a decrease of the proof mass displacement, and therefore $\eta$.
Higher input acceleration leads to a larger value of saturation voltage $V_\mathrm{sat}$. For instance, $A=0.25$ g gives $V_\mathrm{sat} = 29.36$ V while that of $A=0.15$ g is $V_\mathrm{sat} = 22.32$ V.

If the variable capacitor $C_1$ is replaced by a sinusoidal input voltage source $v(t) = V_\mathrm{in} \cos (\omega t)$ and all losses in diodes are neglected, the value of $V_\mathrm{out}$ is supposed to be two times higher $V_\mathrm{C_\mathrm{b1}}$ and therefore $V_\mathrm{out} = 4 V_\mathrm{in}$. However, the relation becomes much more complicated when the input source is the actual voltage generated by MEMS energy harvester, i.e. $V_\mathrm{C_1}$, since this voltage is now dependent on the mechanical vibration and may have time-varying amplitude. For example, the ratio of $V_\mathrm{out}/V_{C_\mathrm{b1}}$ are 1.61 and 1.58 for $A=0.25$ g and $A=0.15$ g respectively. Simulation results also show that $A=0.12$ g is the minimum required acceleration for efficient operation of the circuit, at the resonant frequency $f=f_0$.

\begin{figure}[!t]
	\centering
	\includegraphics[width=0.385\textwidth]{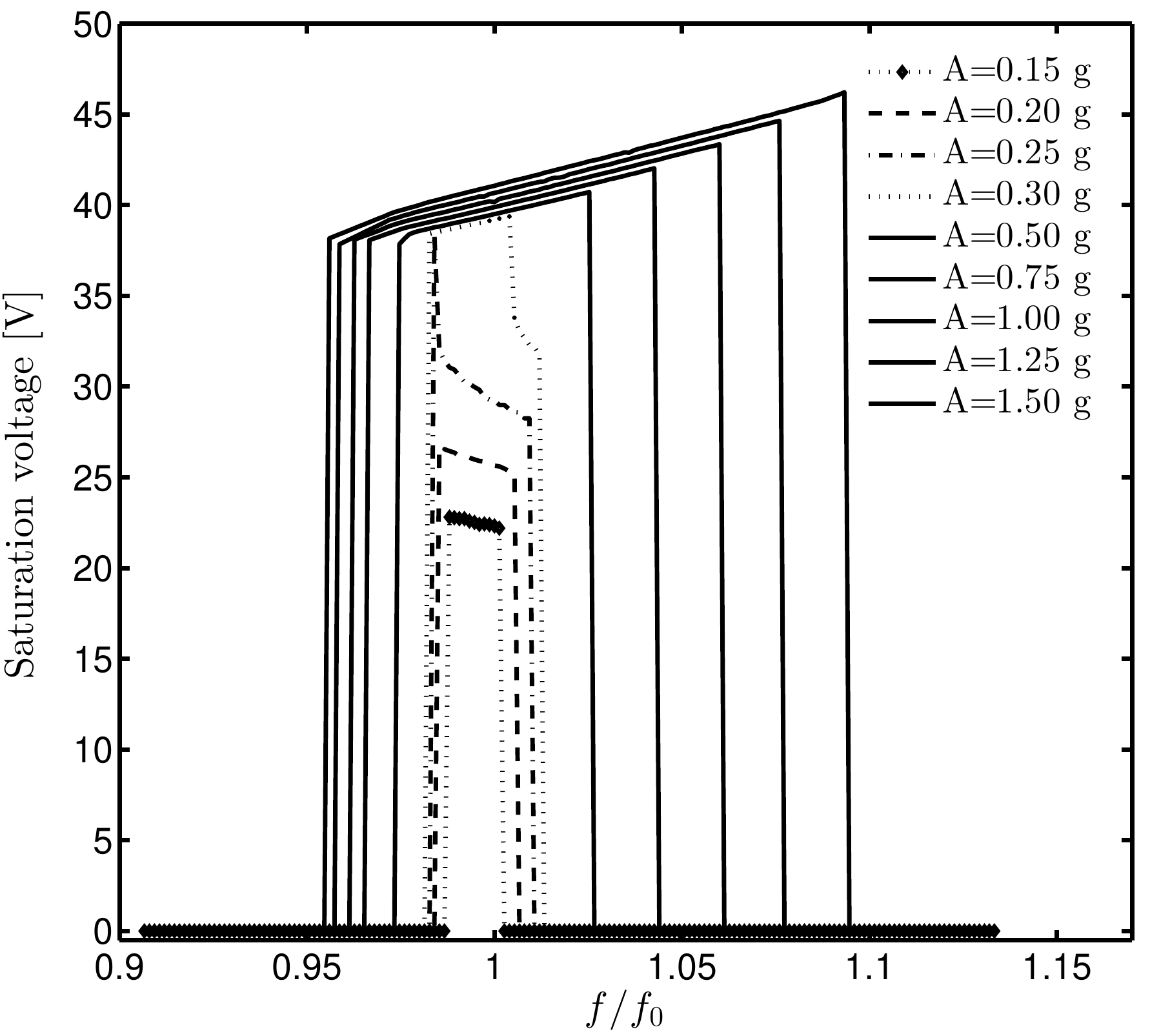}
	\caption{Variation of the saturation voltage with different acceleration amplitudes and frequencies.}
	\label{fig_Vout_Diff_f}
\end{figure}
Figure \ref{fig_Vout_Diff_f} shows variation of the output voltage at steady state obtained from simulations for each pair of acceleration amplitude and frequency ($A,\, f$). At a particular acceleration, when the vibration frequencies are within a certain range around the mechanical resonance, the system harvests energy and the voltage increases towards saturation. For frequencies outside this range, the necessary condition $\eta \geq 1.55$ is not satisfied due to small proof mass displacement. All the stored energy from precharge of the capacitor will then be dissipated resulting in zero output voltage after a certain time.
It is clear that larger excitation strengths create a wider band of useful drive frequencies. 

\begin{figure}[!thb]
		\centering
		\includegraphics[width=0.385\textwidth]{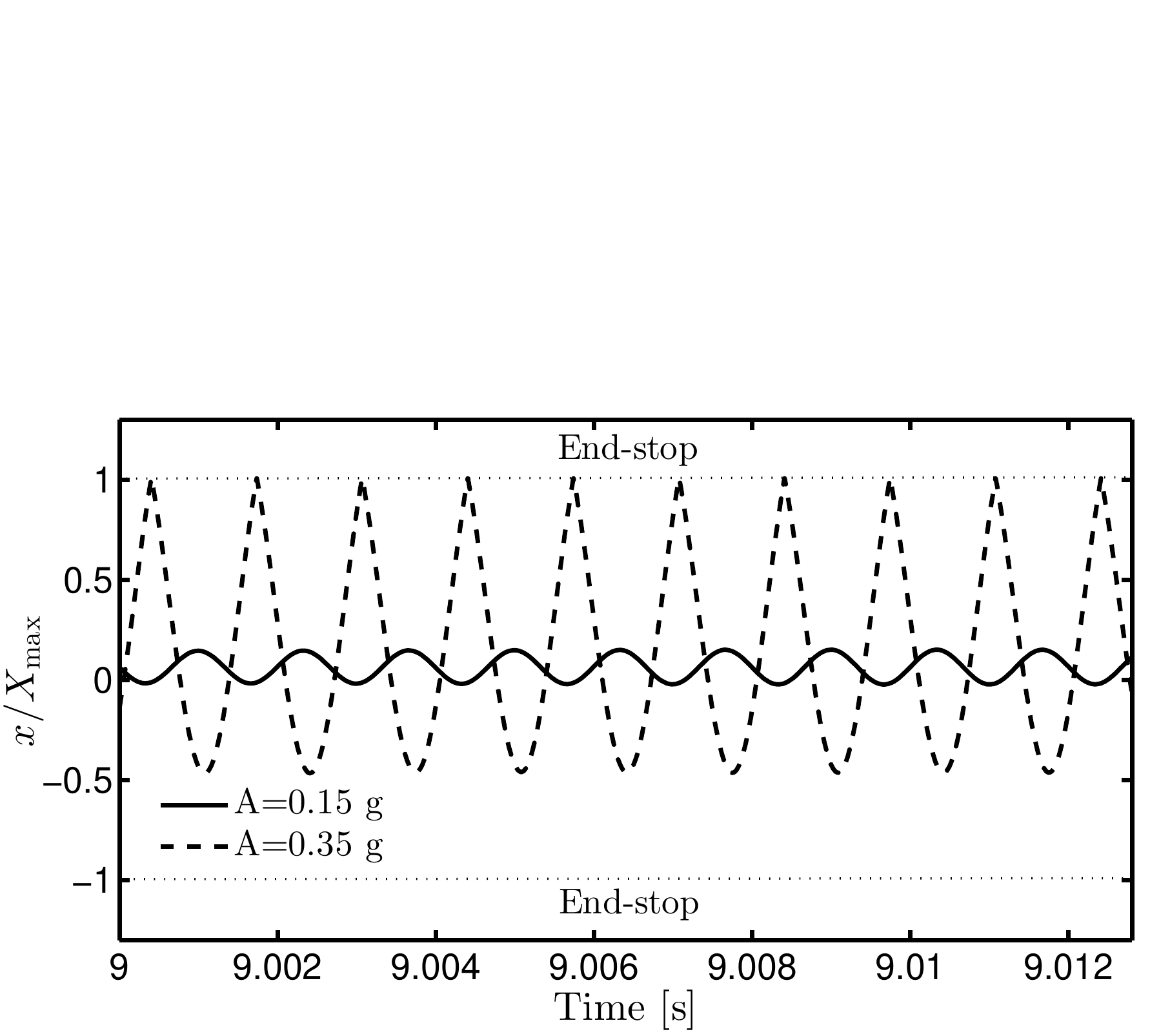}
		\caption{Comparison of the proof mass displacement at steady state for different acceleration amplitudes.}
		\label{fig_x_ss_Diff_A}
\end{figure}
\begin{figure}[!thb]
	\centering
	\includegraphics[width=0.385\textwidth]{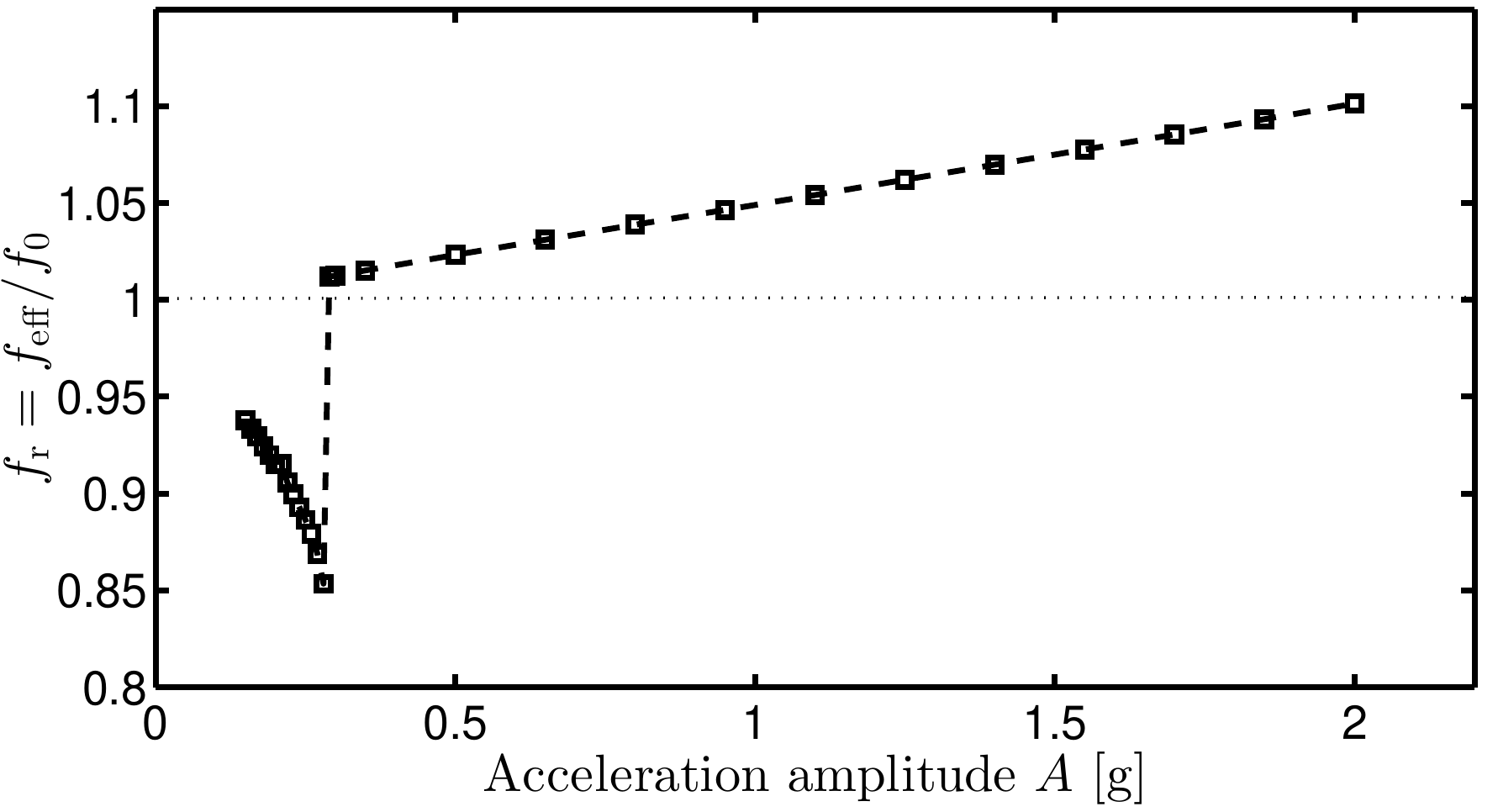}
	\caption{Dependence of the ratio between the effective resonant frequency
		and the resonant frequency on input acceleration.}
	\label{fig_Stiff_ratio}
\end{figure}
In order to understand the characteristics observed in Figure \ref{fig_Vout_Diff_f}, we first consider dynamics of the proof mass at steady state. 
A comparison of normalized signal waveform $x/X_\mathrm{max}$ for different acceleration amplitudes is shown in Figure \ref{fig_x_ss_Diff_A} where $A=0.15$ g and $A=0.35$ g are taken as examples. Further simulation results show that, in the range of $A=0.15$ g$- 0.30$ g, the displacement barely reaches the maximum value $X_\mathrm{max}$ and hence $F_\mathrm{s}=0$. For higher excitation levels, the electrostatic force drives the proof mass toward (closer to) one side of the end-stop and the impact force between them periodically occurs.

We now examine an effective stiffness $K_\mathrm{eff}$ of the anti-phase gap-closing device at steady state, which is given by
\begin{equation}
K_\mathrm{eff} = \Re\mathrm{e}  \left\{  \frac{ \int_{0}^{T} F_\mathrm{ME} \big(t \big) e^{-j \omega t} d\mathrm{t}} {\int_{0}^{T} x \big(t \big) e^{-j \omega t} d\mathrm{t}} \right\},
\end{equation}
where $F_\mathrm{ME} = F_\mathrm{m} + F_\mathrm{e} + F_\mathrm{s}$. The signals of $x \big(t \big)$, $F_\mathrm{m} \big(t \big)$, $F_\mathrm{e} \big(t \big)$ and $F_\mathrm{s} \big(t \big)$ are extracted from simulations. 
The ratio between the effective resonant frequency $f_\mathrm{eff}=\sqrt{K_\mathrm{eff}/m}/2\pi$ and the mechanical resonant frequency is then
\begin{equation}
f_\mathrm{r} =  \frac{f_\mathrm{eff}}{f_0} =  \sqrt{ \frac{K_\mathrm{eff}}{k} }.
\end{equation}
Figure \ref{fig_Stiff_ratio} shows the changes of $f_\mathrm{r}$ versus acceleration amplitude.
In case of $A \in [0.15 \, \mathrm{g}, \, 0.30 \, \mathrm{g}]$, the resonant peak of the transducer is shifted downwards along with the rise of bias voltage due to the softening effect of the nonlinear electrostatic force. Therefore, lower input frequencies result in higher voltages. When $A>0.30$ g, the impact force dominates the electrostatic force and performs the hardening nonlinearity. The resonant peak thus increases in accordance with acceleration amplitudes and the output responses tend to increase with higher frequencies.

Our previous work showed that the saturation voltage of the Bennet's doubler circuit does not depend on the initial bias $V_0$, and can be optimized by increasing the mechanical quality factor and adjusting stiffness \cite{TruongSensors2017a}. These statements are still valid with the voltage multiplier configuration. In this paper, we continue exploring other aspects of the circuit such as effect of the biasing capacitor on start-up time (i.e., the time duration from the beginning to when the output voltage is saturated) and dependence of the ripple voltage on the storage capacitor. It is clearly that minimum start-up time and minimum ripple voltage would be desirable.

\begin{figure}[!tbp]
	\centering
	\includegraphics[width=0.385\textwidth]{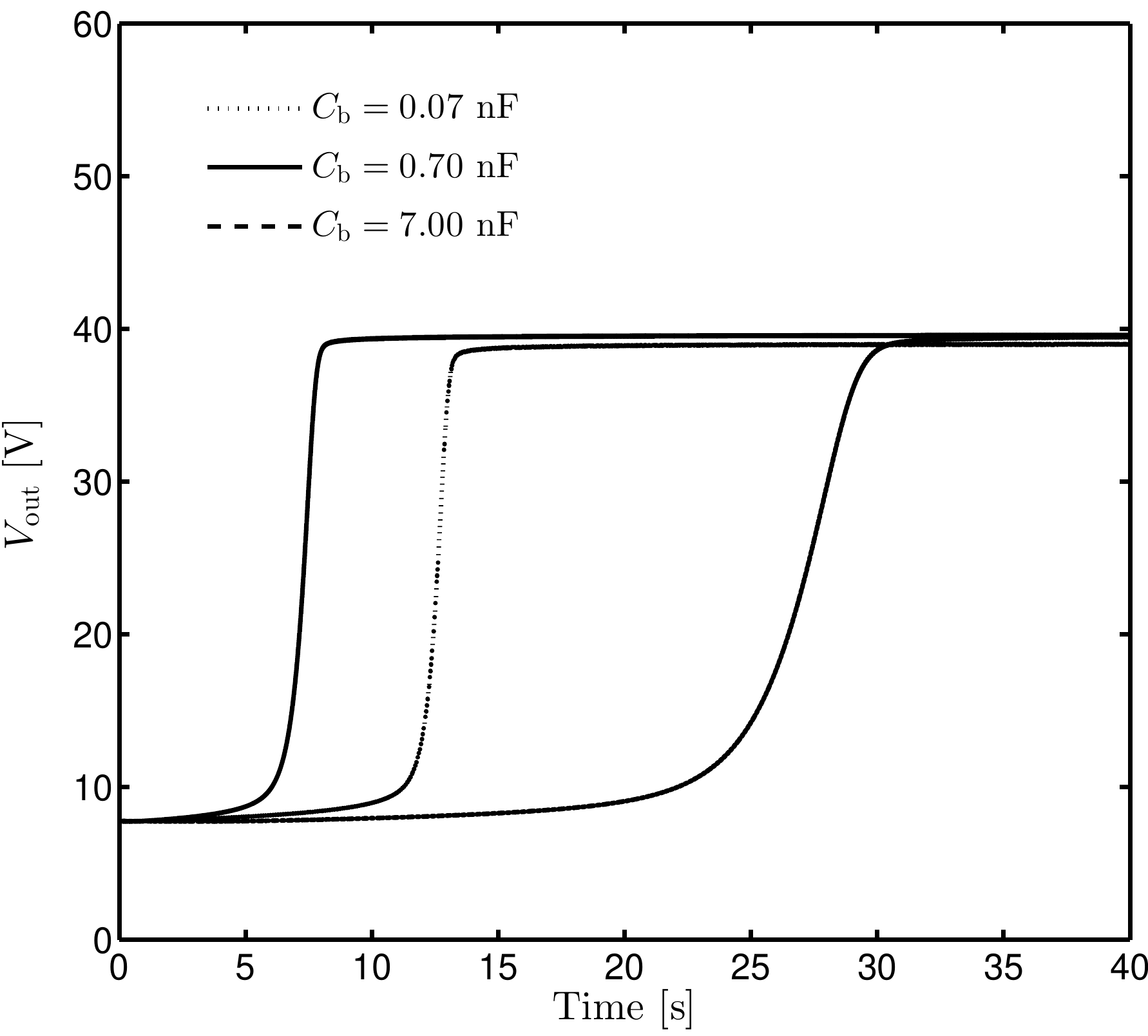}
	\caption{The time evolution of the output voltage $V_\mathrm{out}$ across $C_\mathrm{s}$ for different values of $C_\mathrm{b}$.}
	\label{fig_Vout_Diff_Cb}
\end{figure}
\begin{figure}[!tbp]
	\centering
	\includegraphics[width=0.385\textwidth]{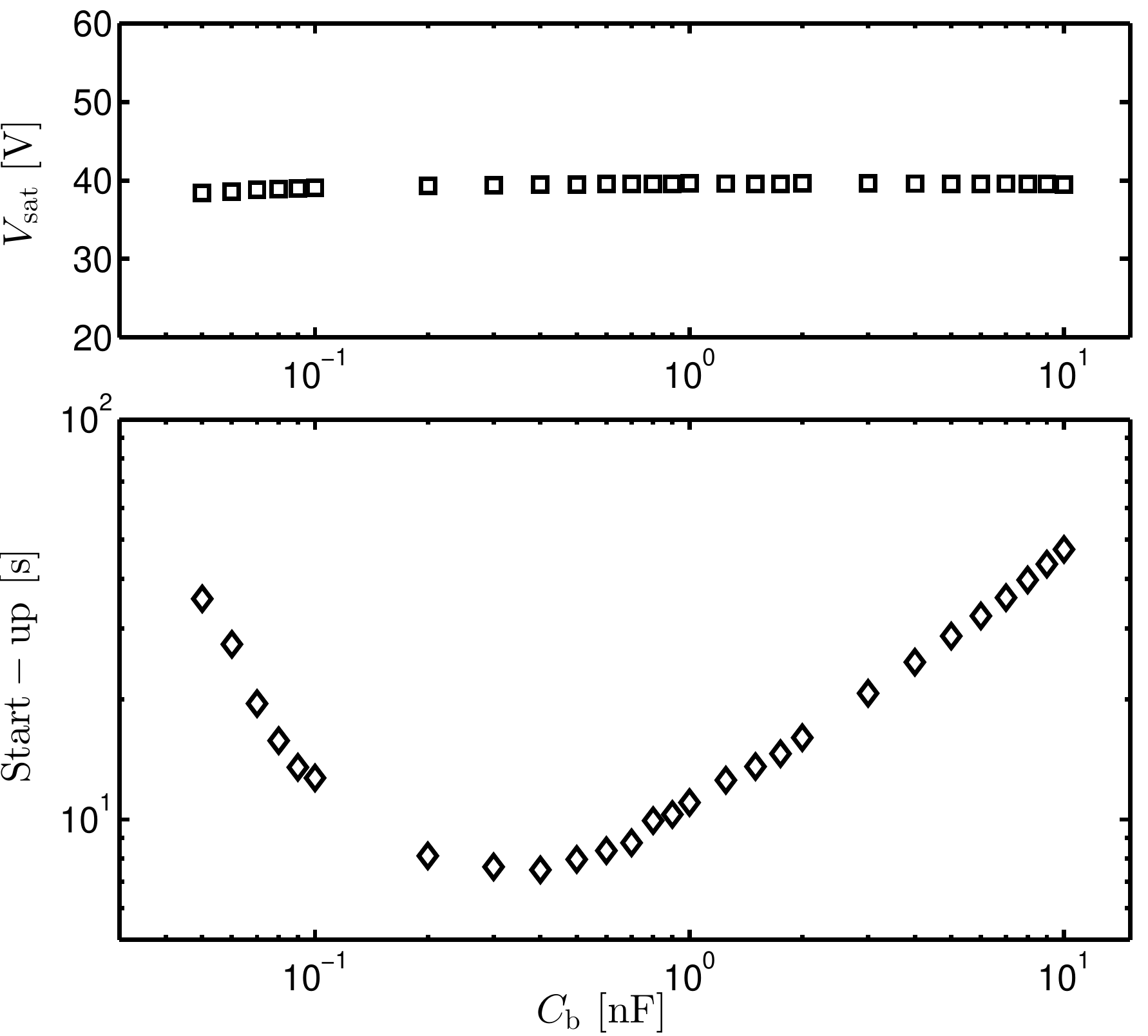}
	\caption{Start-up time for different values of $C_\mathrm{b}$.}
	\label{fig_StUp_Diff_Cb}
\end{figure}
For simplification, all fixed capacitances are chosen equal $C_\mathrm{b1} = C_\mathrm{b2} = C_\mathrm{b}$. Figure \ref{fig_Vout_Diff_Cb} depicts the effect of $C_\mathrm{b}$ on the evolution of output voltage over time. The biasing capacitance $C_\mathrm{b}$ almost does not affect the saturation voltage $V_\mathrm{sat}$, but has strong influence to the required time duration until $V_\mathrm{sat}$ is achieved. Therefore, the start-up time can be minimized with respect to $C_\mathrm{b}$ so that the circuit can produce the same expected amount of the DC voltage faster.
The start-up time used to distinguish between the transient and steady states is determined once the output voltage reaches $98\%$ of the saturation voltage. Simulation results obtained for a wide range of $C_\mathrm{b}$ are shown in Figure \ref{fig_StUp_Diff_Cb}. The saturation voltage can be considered unchanged, while the shortest start-up time $t_\mathrm{min} \approx 7.5$ s is found around $C_\mathrm{b} = 0.4$ nF. Larger capacitances result in longer changing time for each multiplier-stage and thus slow down the start-up, but too small ones even increase the time sharply due to the undesired dynamic loading condition. From another perspective, $C_\mathrm{b}$ has a significant impact on the cycle-to-cycle multiplication factor $z$ of the output voltage. The argument is similar to the one that was used to explain the influence of the storage capacitor on $z$ for the Bennet's doubler circuit \cite{deQueiroz2011, TruongSensors2017a}.

\begin{figure}[!tbp]
	\centering
	\includegraphics[width=0.385\textwidth]{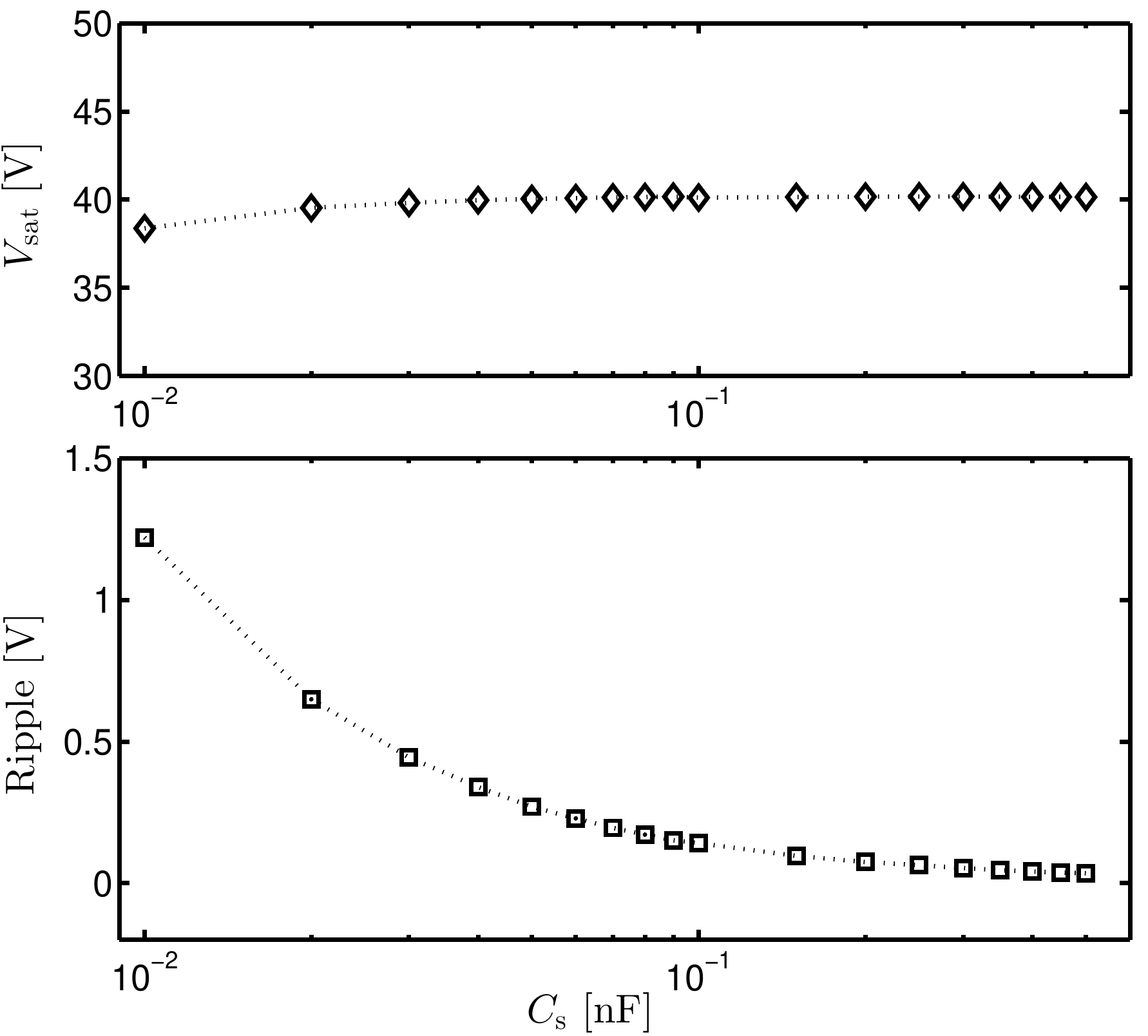}
	\caption{The output DC voltage and the corresponding ripple with different values of $C_\mathrm{s}$.}
	\label{fig_Vout_Rippler}
\end{figure}
Simulation results for the voltage ripple and the corresponding saturation voltage versus the storage capacitances are shown in Figure \ref{fig_Vout_Rippler}. The ripple is determined as $( V_\mathrm{max} - V_\mathrm{min} ) / 2$, where $V_\mathrm{max}$ and $V_\mathrm{min}$ are the highest and lowest peaks of the output voltage at steady state. Meanwhile, the saturation voltage is $V_\mathrm{sat} \approx (V_\mathrm{max} + V_\mathrm{min})/2 $. The ripple is found to be about 35 mV when $C_\mathrm{s}=0.5$ nF and increase with decrease of $C_\mathrm{s}$. $V_\mathrm{sat}$ is slightly reduced when $C_\mathrm{s} < 50$ pF is used.
Since our objective is to obtain a DC output voltage, it is desirable to choose as large a value of the capacitance $C_\mathrm{s}$ as possible in order to minimize the ripple. However, the analysis presented in Figure \ref{fig_StUp_Diff_Cb} also holds for $C_\mathrm{s}$, meaning that larger capacitances result in longer start-up time. Therefore, a good trade-off is needed here to keep both the ripple and start-up time at reasonable levels that are dependent on particular requirements of the output specifications.

At the input acceleration $A=0.12$ g and the drive frequency $f = f_0$, simulations obtained for the low-losses diode BAS716 show that $V_0=6$ V is the minimum required initial bias for efficient operation of the circuit, so-called critical voltage $V_\mathrm{cr}$.
In previous work we proved that the use of a MOSFET connected in a diode-configuration as an alternative to the $p^+n$ diode could significantly reduce $V_\mathrm{cr}$. Moreover, the diode-connected MOSFET is also a convenient option for integrated circuits. With the same MOSFET parameters given in \cite{TruongSensors2017a}, the value of $V_\mathrm{cr}$ can be further lowered to $1.9$ V.

\section{Generalized N-stage voltage multiplier}

\begin{figure}[!htbp]
	\centering
	\includegraphics[width=0.4\textwidth]{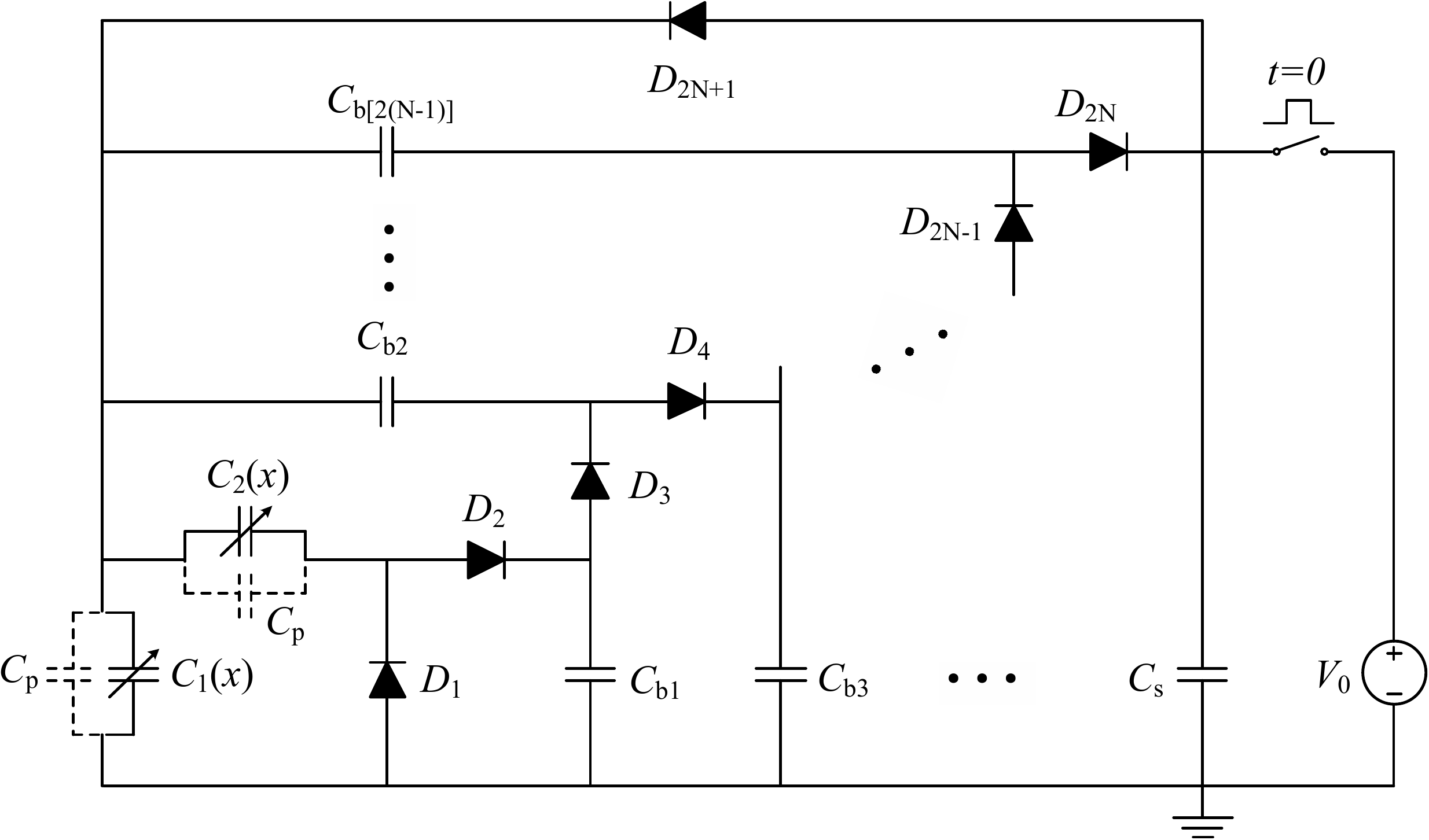}
	\caption{Multi-stage voltage doubler circuit - Configuration (I).}
	\label{fig_Gen_Cir}
\end{figure}
The introduced circuit can be generalized to an N-stage voltage doubler as depicted in Figure \ref{fig_Gen_Cir}, where N is an integer and $N \geq 3$. Each intermediate stage $j \in [2,N-1]$ includes two biasing capacitors $C_\mathrm{b[2(j-1)]}$, $C_\mathrm{b[2(j-1)+1]}$ and two rectifier diodes $D_\mathrm{2j-1}$, $D_\mathrm{2j}$. The fly-back diode $D_\mathrm{2N+1}$ is connected between the storage capacitor $C_\mathrm{s}$ of the final stage and the variable capacitor $C_1$. $D_\mathrm{2N+1}$ plays exactly the same role as $D_5$ does in Figure \ref{fig_Model}.

\begin{figure}[!htbp]
	\centering
	\includegraphics[width=0.35\textwidth]{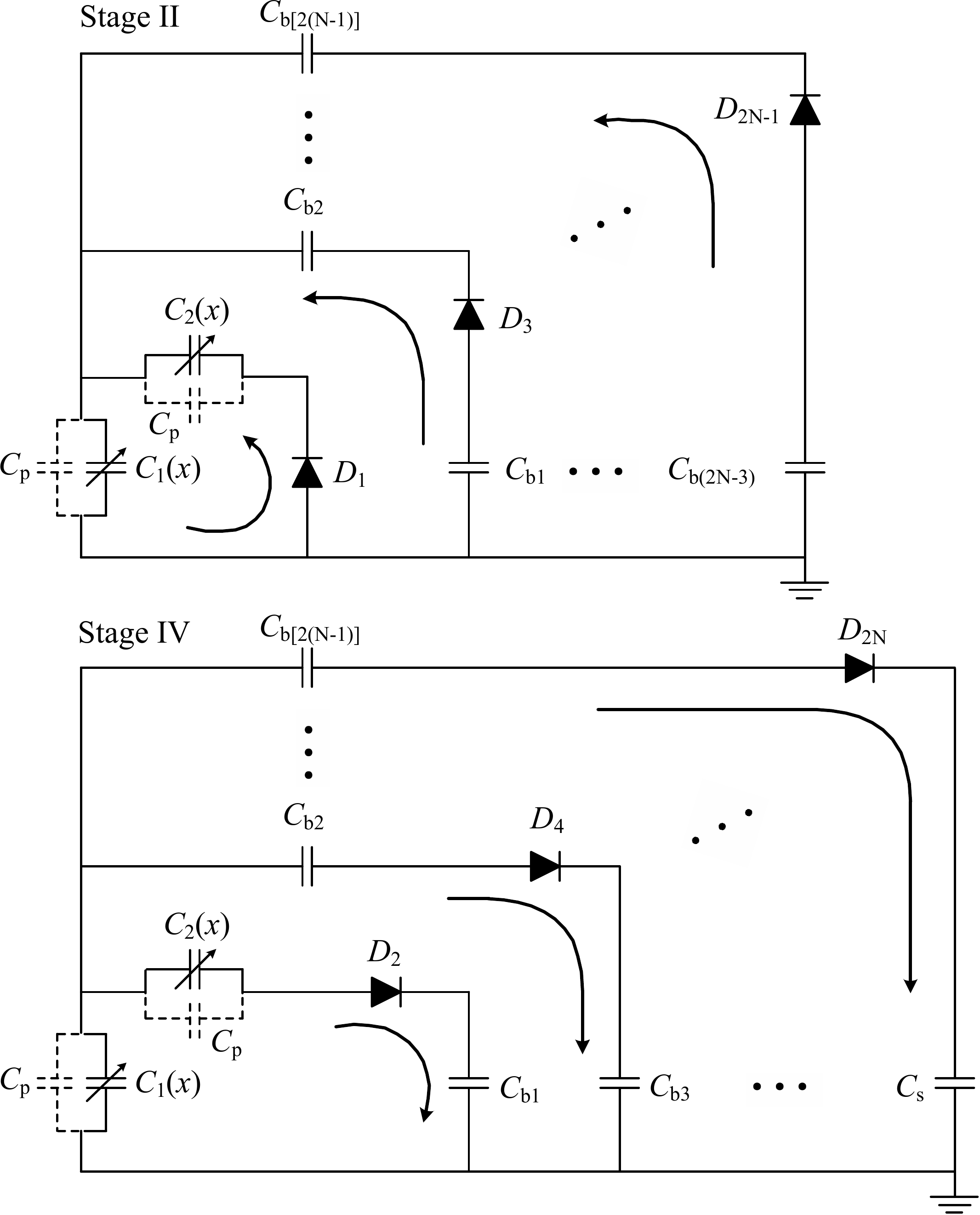}
	\caption{Operation of multi-stage voltage doubler circuit: stage II and IV.}
	\label{fig_Gen_Cir_Operation}
\end{figure}
Despite of complex dynamics of the voltage multiplier configuration, we can also simplify and divide its operation over one cycle into four sequence stages that are similar to the Section \ref{Two-stage}. Since all diodes $D_\mathrm{i} \, (i \in [1,\, 2N+1])$ are blocked in the first and the third stages, we consider two main topologies regarding directions of the parallel currents across diodes and capacitors. Both are represented in Figure \ref{fig_Gen_Cir_Operation}. Though diode $D_\mathrm{2N+1}$ is required for initially charging $C_1$, its conduction is insignificant during operation of the harvesting system, and therefore is disregarded. In the first topology, only $D_\mathrm{2j-1} \ (j \in [1, \, N])$ are conducting. Charges are pumped from capacitor $C_1$ and $C_\mathrm{b[2(k-1)-1]}$ into $C_2$ and $C_\mathrm{b[2(k-1)]} \ (k \in [2,\, N])$ respectively. This stage stops when the proof mass displacement reaches the maximum value $x=X_\mathrm{max}$. The second topology corresponds to discharge of $C_2$ and $C_\mathrm{b[2(k-1)]}$ into $C_\mathrm{b[2(k-1)-1]}$ and $C_\mathrm{s}$ due to conduction of diodes $D_\mathrm{2j}$. Until $x=-X_\mathrm{max}$, all $D_\mathrm{i}$ are blocked again and a new cycle starts. Observing that the currents flow in parallel branches in the conducting regime and the doubler stages are connected in series, such a voltage multiplier configuration can be classified as a parallel-series charge-pump or parallel-series switched/diode-capacitor network.

\begin{figure}[!t]
	\centering
	\includegraphics[width=0.35\textwidth]{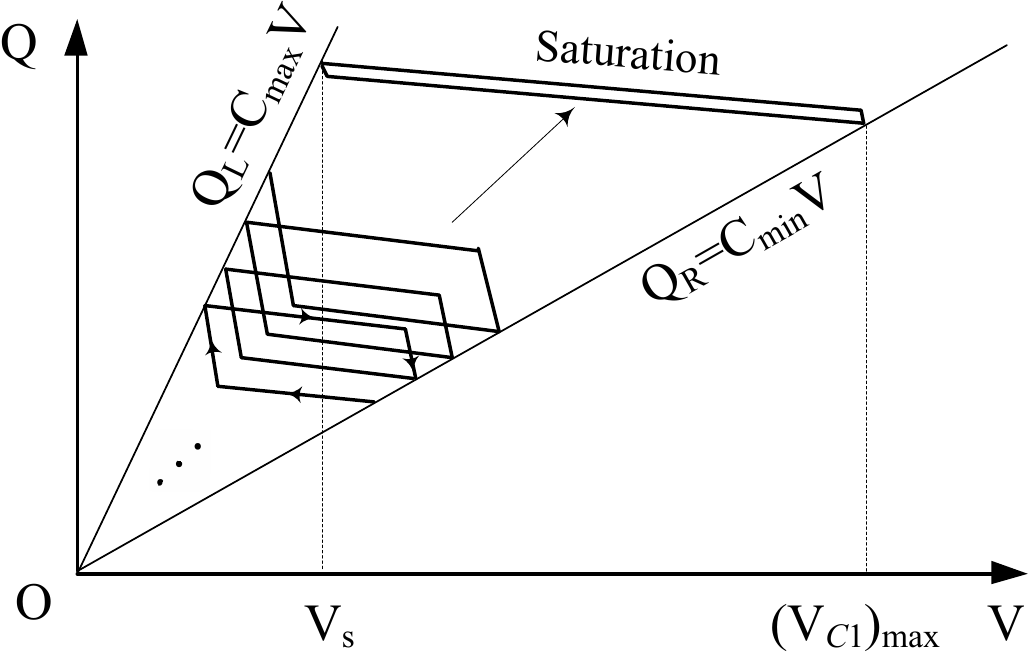}
	\caption{Succession of charge-voltage cycles of $C_1$}
	\label{fig_QV}
\end{figure}
Based on the dynamic simulation, the operation of capacitor $C_1 (x)$ can be approximated by a Q-V cycle depicted in Figure \ref{fig_QV}. In the left and the right sides, the Q-V diagram is bounded by the lines corresponding to maximum and minimum values of the variable capacitance: $Q_\mathrm{L}=C_\mathrm{max} V$ and $Q_\mathrm{R}=C_\mathrm{min} V$.
The area of each parallelogram is roughly equal to the harvested energy per cycle \cite{Galayko2008}, and is modified over cycles in a complicated manner. It is very small at the beginning and gradually increases in the conversion regime. When the output voltage across the storage capacitor $C_\mathrm{s}$ approaches the saturation voltage $V_\mathrm{out} \approx V_\mathrm{sat}$, the shape of the Q-V cycle becomes thinner. At $n \longrightarrow \infty$, it degenerates to the line, meaning that the mechanical energy is no longer captured. Here, $n$ is the number of cycles. The similarity was observed from the Q-V cycle represented for the charge-pump circuit with resistive fly-back \cite{Florentino2011, O'Riordan2015}.

\begin{figure}[!htbp]
	\centering
	\includegraphics[width=0.385\textwidth]{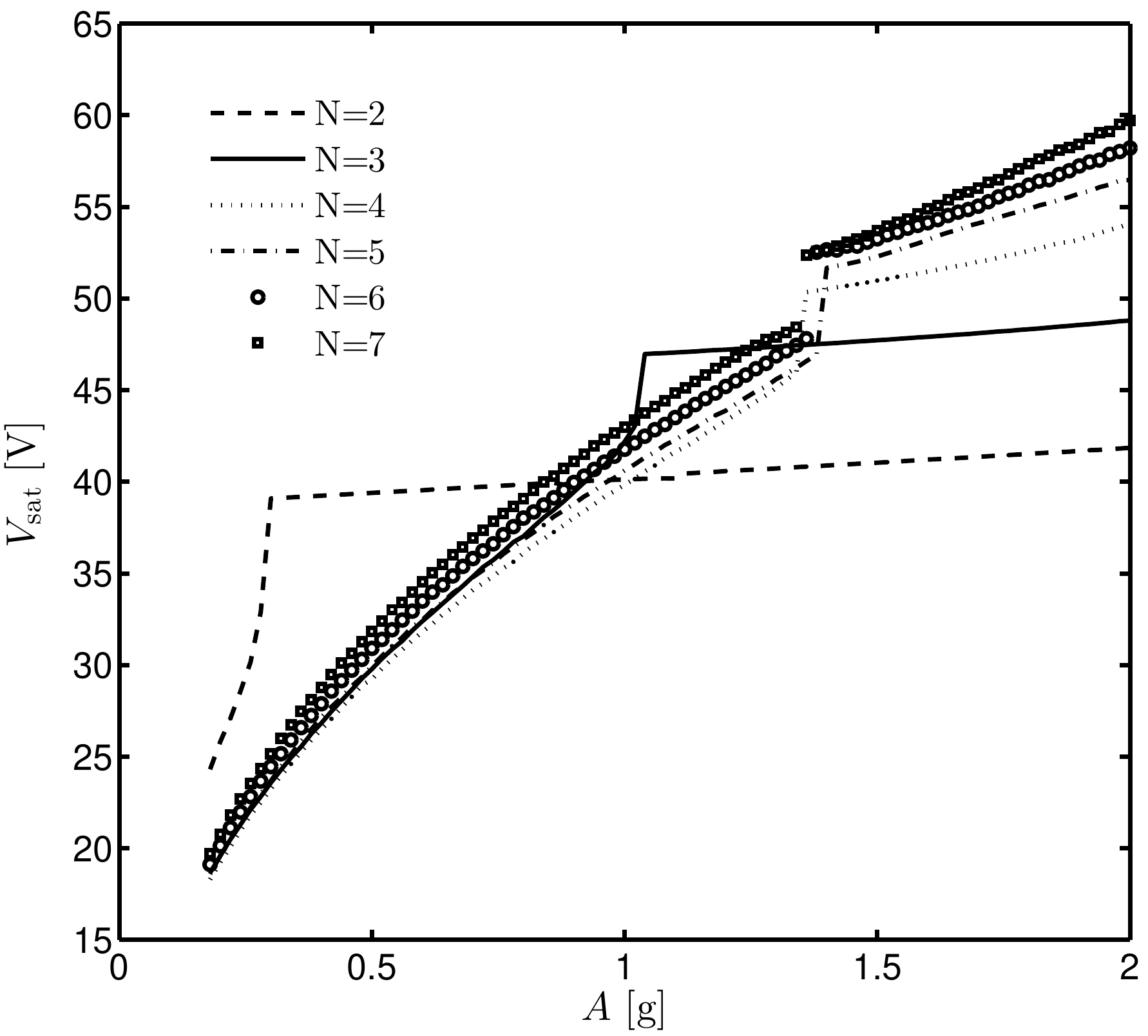}
	\caption{Output voltage at saturation versus input acceleration with different value of the number of stages $N$.}
	\label{fig_Vout_A_N}
\end{figure}
Figure \ref{fig_Vout_A_N} shows the saturation voltage of the circuit versus the input accelerations for various number of stages $N$.
\begin{figure}[!tbp]
	\centering
	\includegraphics[width=0.385\textwidth]{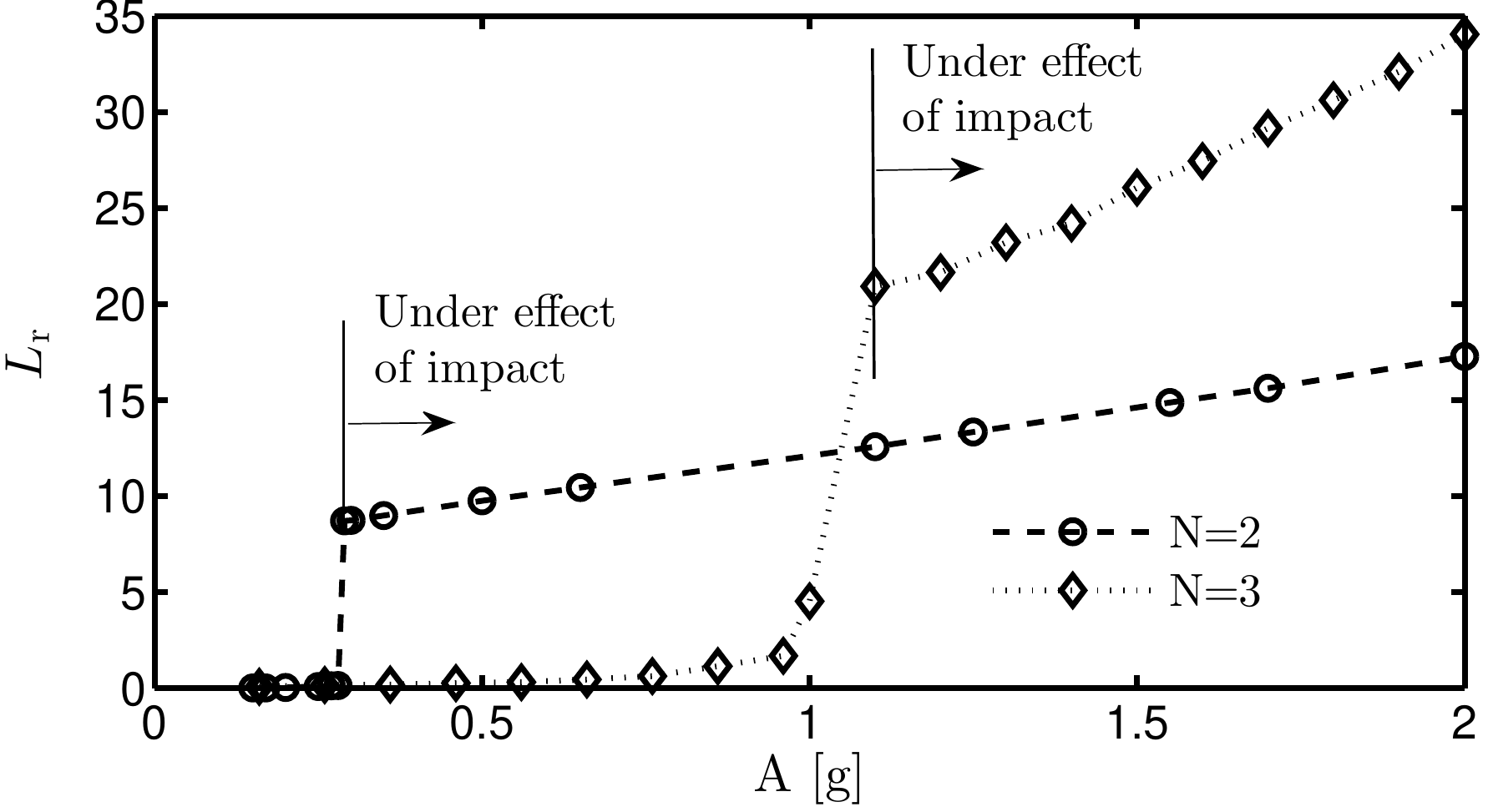}
	\caption{Dependence of the ratio between average power lost in both electrical damping and impact force and mechanical damping on input acceleration amplitude}
	\label{Fig:Lr_A}
\end{figure}
Considering the two-stage voltage doubler $N=2$, there is a rapid increase of the saturation voltage corresponding to increase of $A \in [0.15 \, \mathrm{g}, \, 0.30 \, \mathrm{g}]$. However, higher acceleration amplitudes result in very slight rise of $V_\mathrm{sat}$. Similar behaviors are also observed for $N\geq3$. To understand the reason for this phenomenon, we consider the relation between the work done against both electrostatic and impact forces and the energy lost in mechanical damping per cycle, which is characterized by the ratio
\begin{equation}
L_\mathrm{r} = \frac{\int_{0}^{T} [F_\mathrm{e}  \big(t \big) + F_\mathrm{s}  \big(t \big)]\; \dot{x} \big(t \big) \mathrm{d}t}{\int_{0}^{T} B(t)\; \dot{x}(t)^2\, \mathrm{d}t} 
\end{equation}
where the electrostatic force $F_\mathrm{e} \big(t \big)$, the impact force $F_\mathrm{s} \big(t \big)$, the time-dependent damping $B \big(t \big)$ and the proof mass velocity $\dot{x} \big(t \big)$ are extracted from simulations at steady state. The variation of $L_\mathrm{r}$ versus acceleration amplitude for $N=2$ and $N=3$ are depicted in Figure \ref{Fig:Lr_A} as examples. Sufficiently high accelerations (i.e., $A > 0.3$ g with $N=2$ and $A > 1.0$ g with $N=3$) result in collisions between the proof mass and the end-stops even at steady state. The coincidence between the figure \ref{fig_Vout_A_N} and \ref{Fig:Lr_A} shows that the energy losses due to impacts make the saturation voltage increase very little with $A$.

In addition, there is a range of input excitation that N-stage multiplier circuit with $N\geq3$ performs worse than the case of $N=2$. For instance, the 3-stage voltage multiplier only gives more benefit on the saturation voltage than the 2-stage configuration when $A$ is equal or greater than a certain value called critical acceleration, in detail $A \geq A_\mathrm{cr}=0.94$ g. It is also observed that $A_\mathrm{cr}$ decreases with increase of $N$, e.g. $A_\mathrm{cr} \approx 0.84$ g with $N=7$. All those behaviors are due to effect of the strong nonlinear electrostatic force.


\begin{figure}[!t]
	\centering
	\includegraphics[width=0.385\textwidth]{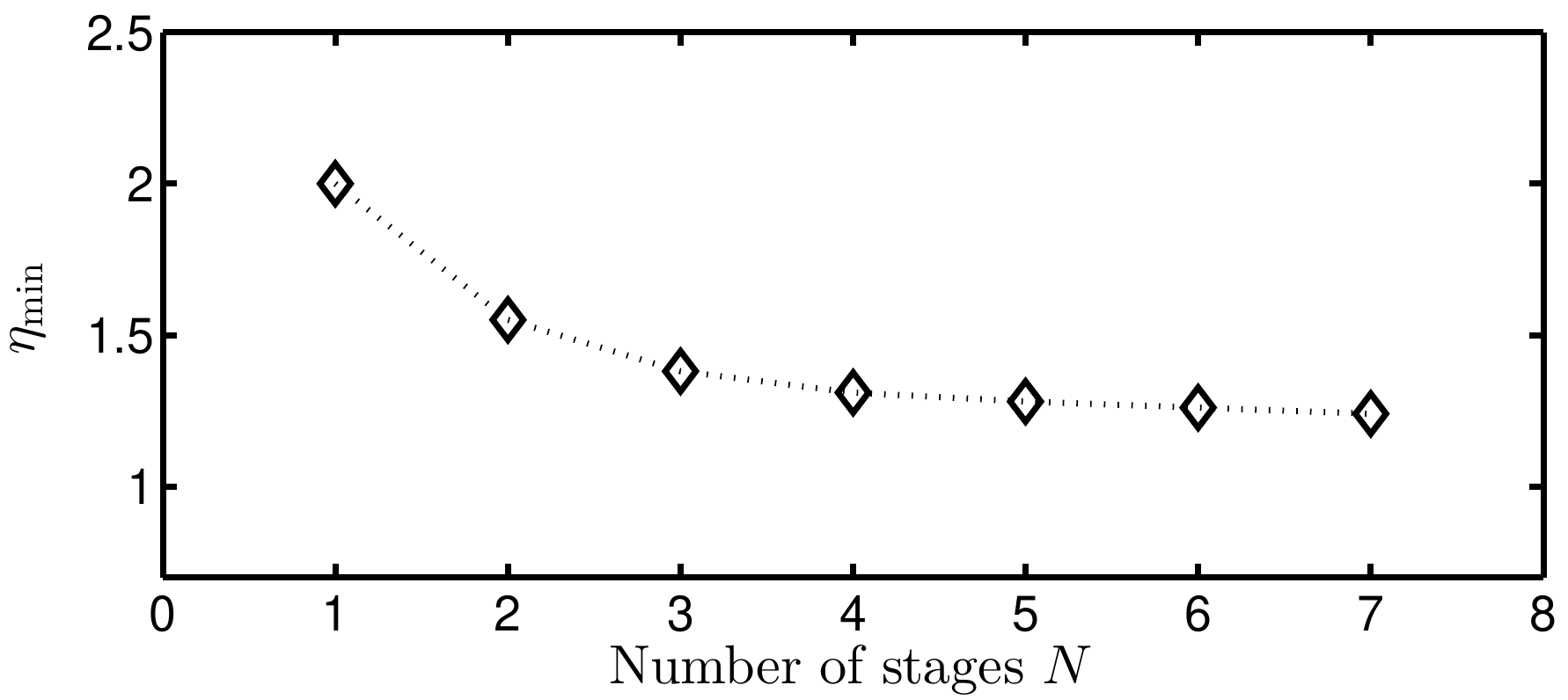}
	\caption{Minimum value of the capacitance ratio $\eta_\mathrm{min}$ versus number of stages $N$.}
	\label{fig_Vout_N}
\end{figure}
One important criteria to evaluate and highlight the performance of the voltage multiplier is its capability to work with micro-scale energy harvesters or low power systems.
The minimum value of the capacitance ratio $\eta_\mathrm{min}$ for different $N$ is given in Figure \ref{fig_Vout_N}, with input acceleration $A=0.5$ g. In the ideal case, the prerequisite condition for operation of the circuit is that $\eta$ must be greater than the threshold $\eta ^ \ast  = \frac{N+1}{N}$ \cite{deQueiroz2016}. It is interesting to note that $\eta ^ \ast$ and $\eta_\mathrm{min}$ can be decreased by increasing $N$. When the electrical losses are taken into account, $\eta_\mathrm{min}$ is slightly higher than $\eta ^ \ast$. Even so, the presented configuration is a promising solution to overcome the challenging obstacle of MEMS devices which typically have $\eta < 2$ due to the limitation of device size, micro-fabrication process and presence of parasitic capacitance.

Based on the analyses above, it can be concluded that with the use of the anti-phase gap-closing devices, 2-stage voltage multiplier is a better option for transducers with higher ratio $\eta$ operating at low input acceleration amplitudes while 3- or more-stage ones are prefer in the opposite cases.

\section{Comparison of different topologies of the voltage multiplier}

Due to mentioned problems, there are only a few circuits for which successful realizations, at least in theory, have been reported so far. A topology proposed by Lefeuvre \textit{et al.} \cite{Lefeuvre2014} was given that the minimum necessary ratio $\eta$ for its operation can be lower than 2. The argument is based on theoretical analysis of the rectangular Q-V cycle alone, there are no concrete evidences in both dynamic simulations and measurements. Instead, the experimental validations were carried out using a single rotating variable capacitor driven by a DC motor, with nearly $C_\mathrm{max}=45$ pF and $C_\mathrm{min}=155$ pF. From the same group, a similar contribution was made in \cite{Wei2016}, where $C_\mathrm{min}=25$ pF and $C_\mathrm{max}=125$ pF. It is clearly that the ratio $C_\mathrm{max}/C_\mathrm{min}$ in both cases is much lager than 2. 
In the same manner, the work of Karami \textit{et al.} \cite{Karami2017} presented a series-parallel charge-pump conditioning circuits which are generalized from the Bennet's doubler. However, this contribution is limited to an electrical domain study only. One may concern about the effect of the electromechanical coupling on the full dynamics of the energy harvesters.

In this paper, we choose to compare our voltage multiplier configuration with the one in \cite{Lefeuvre2014}.
The series-parallel voltage doubler \cite{Karami2017} is explored and discussed afterward, due to some limitations that are revealed by our simulation results.

\begin{figure}[!htbp]
	\centering
	\includegraphics[width=0.40\textwidth]{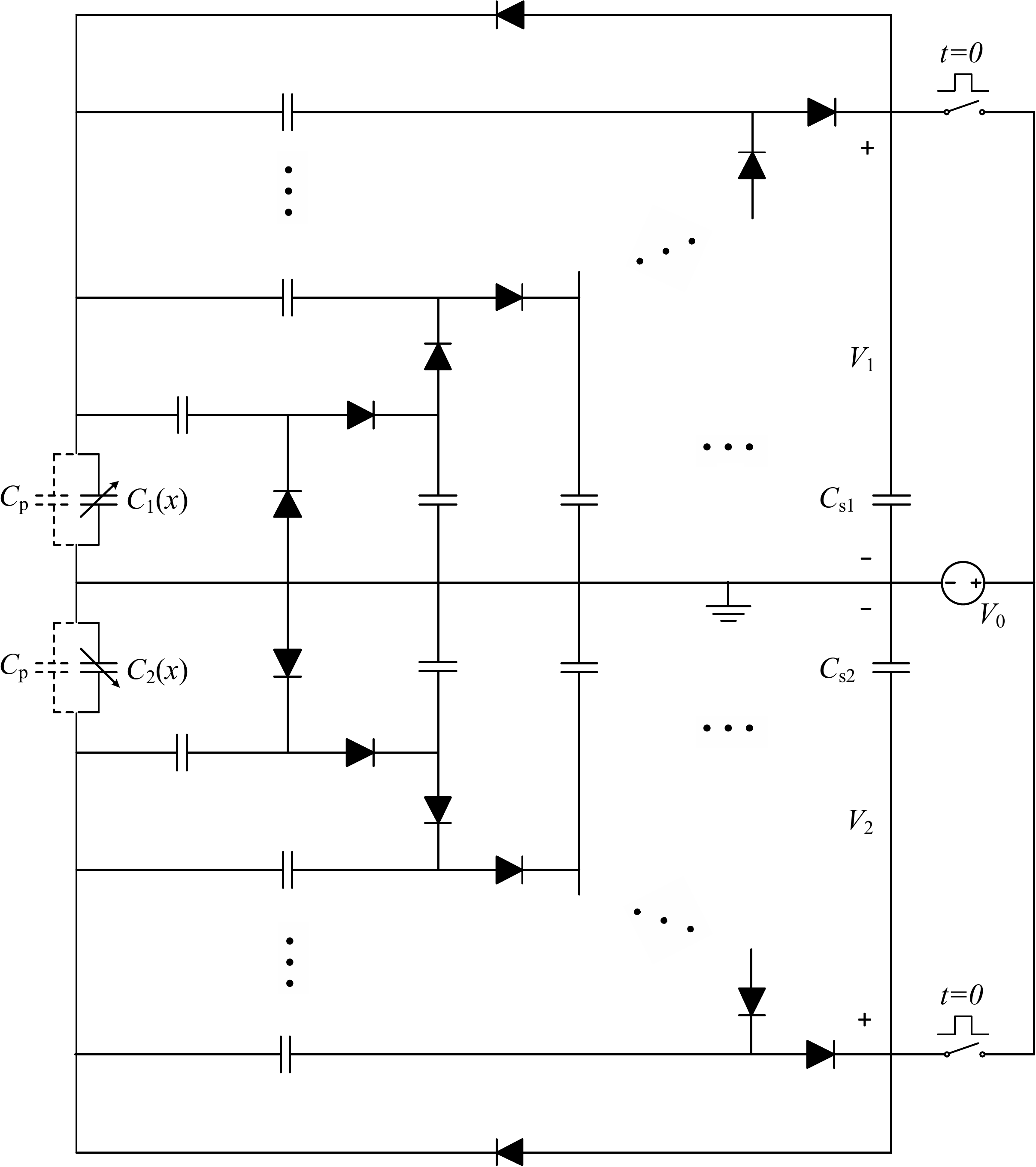}
	\caption{Symmetric multi-stage voltage doubler circuit - Configuration (II).}
	\label{fig_Multi_N_Symmetric}
\end{figure}
\begin{figure}[ht!]
	\begin{center}
		\subfigure[Congifuration (III)]{%
			\label{fig:Lefeuvre_1}
			\includegraphics[width=0.4\textwidth]{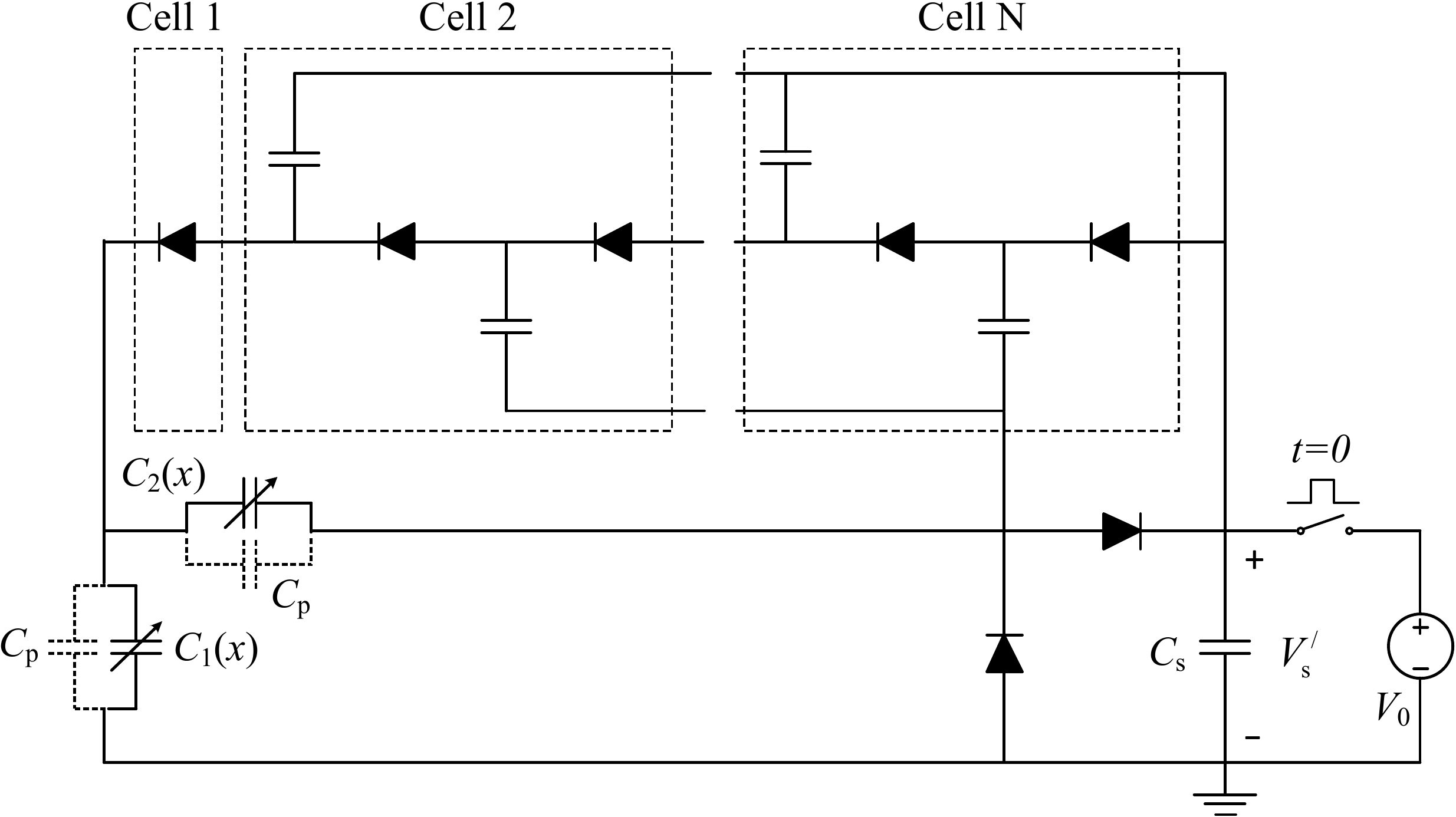}
		}\\
		
		\subfigure[Congifuration (IV)]{%
			\label{fig:Lefeuvre_2}
			\includegraphics[width=0.4\textwidth]{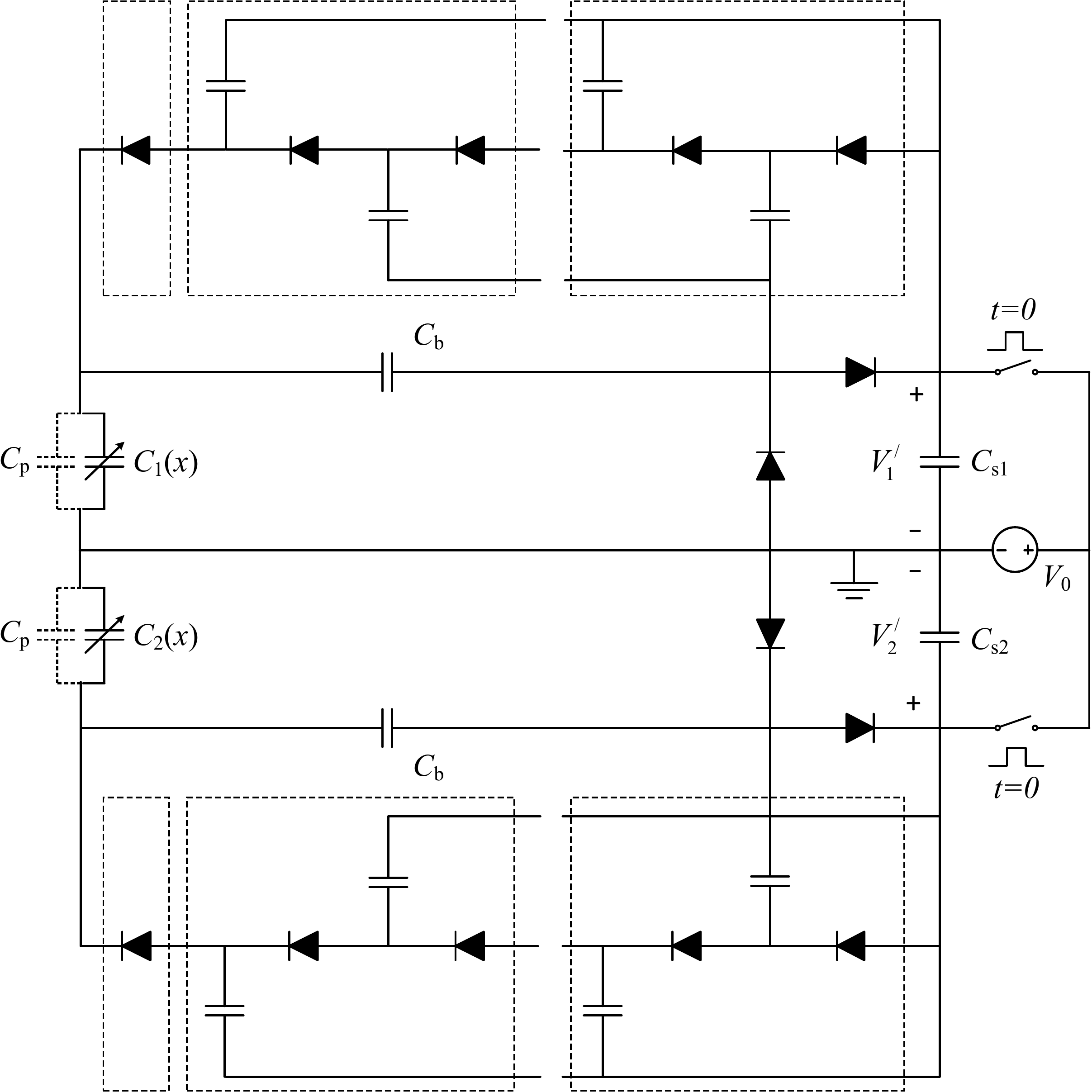}
		}
	\end{center}
	\caption{Schematic diagram of the self-biased interface circuits \cite{Lefeuvre2014}.
	}%
	\label{fig:Lefeuvre_both}
\end{figure}
Besides the circuit shown in Figure \ref{fig_Gen_Cir}, an anti-phase structure may provide us another configuration where each transducer produces a single output voltage, see Figure \ref{fig_Multi_N_Symmetric}. The insignificant anti-phase AC signals referred as ripple voltages are neglected. The two output voltages can be considered equal $V_1 = V_2$, which is the symmetric characteristics of the circuit. Similarly, we can divide the circuit configurations proposed by Lefeuvre \textit{et al.} into two varieties that are depicted in Figure \ref{fig:Lefeuvre_both}. Each cell $\overline{2,N}$ is an adaptation of the Greinacher voltage doubler. In configuration (IV) we also get $V_1^{'} = V_2^{'}$. For a fairly comparison, we classify four different configurations into two group: (I)/(III) and (II)/(IV), as each group shares the same number of electrical outputs.

\begin{figure}[!htbp]
	\centering
	\includegraphics[width=0.385\textwidth]{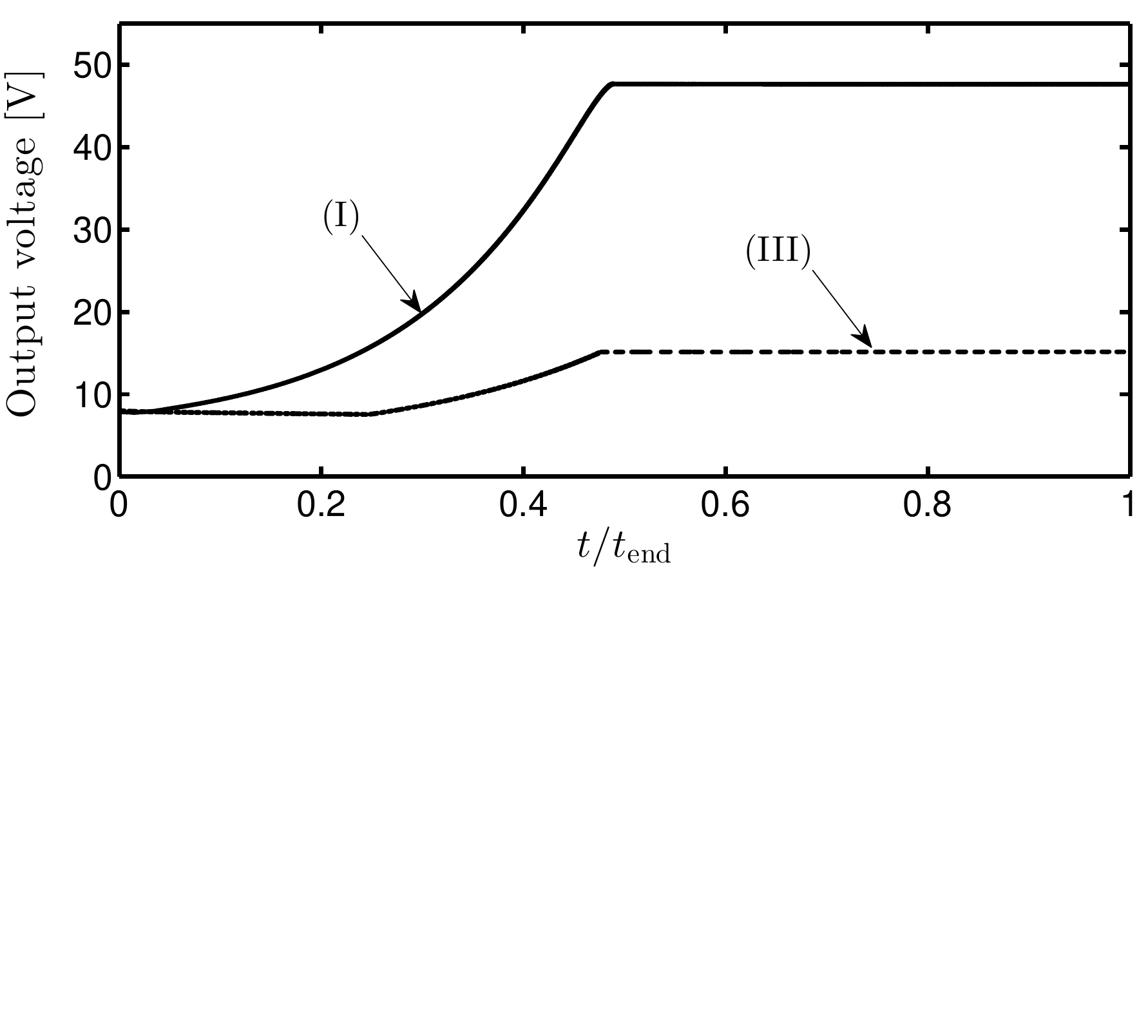}
	\caption{Time evolution of output voltage: configuration (I) and (III).}
	\label{fig_Diff_Configs_1}
\end{figure}
\begin{figure}[!htbp]
	\centering
	\includegraphics[width=0.385\textwidth]{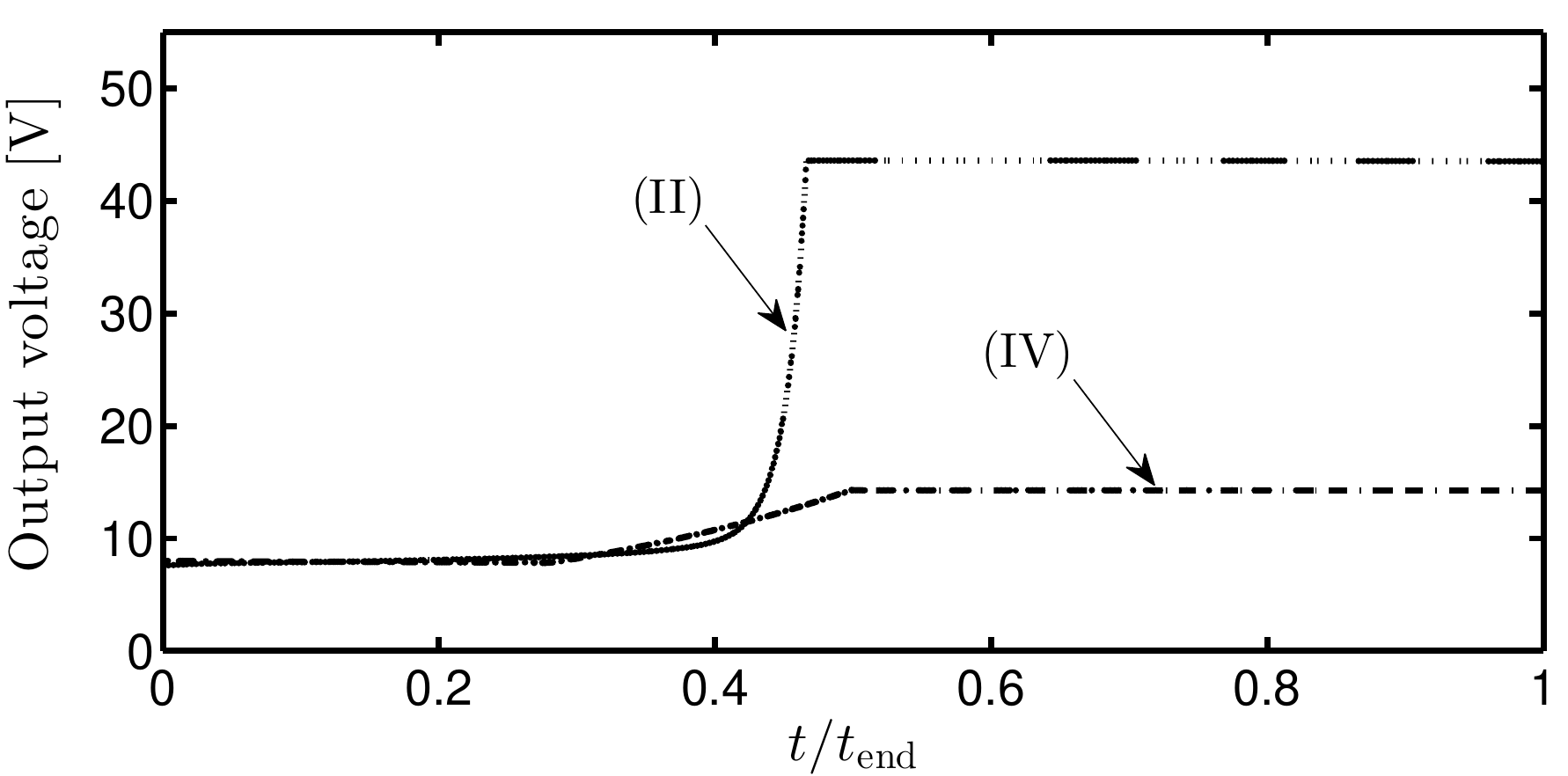}
	\caption{Time evolution of output voltage: configuration (II) and (IV).}
	\label{fig_Diff_Configs_2}
\end{figure}
Figure \ref{fig_Diff_Configs_1} and \ref{fig_Diff_Configs_2} shows the evolution of output voltage for configuration (I)/(III) and (II)/(IV) respectively, with the input acceleration $A=1.5$ g, drive frequency $f=f_0$, initial bias $V_0 = 8$ V and the number of stage $N=3$. It should be noted that $\eta=1.55$ is not enough for efficient operation of both configurations (II) and (IV) with $N=2$. The end-time of simulations are $t_\mathrm{end}^\mathrm{(I)} = 2.8,\, t_\mathrm{end}^\mathrm{(II)} = 65, \, t_\mathrm{end}^\mathrm{(III)} = 1.3, \, t_\mathrm{end}^\mathrm{(IV)} = 4$ seconds. A major disadvantage of the configuration (II) over the others is its long start-up time, which however can be shortened by carefully choosing the biasing/storage capacitor and increasing $V_0$. For example, the start-up time of the configuration (II) reduces from 30.6 s with $V_0=8$ V and $C_\mathrm{b}=0.5$ nF to 7.4 s with $V_0=9$ V and $C_\mathrm{b}=0.2$ nF.
We see that the output voltage of configuration (I) saturates at much higher value than that of configuration (III). The same phenomenon is observed when configuration (II) is compared to configuration (IV).

\begin{figure}[!htbp]
	\centering
	\includegraphics[width=0.385\textwidth]{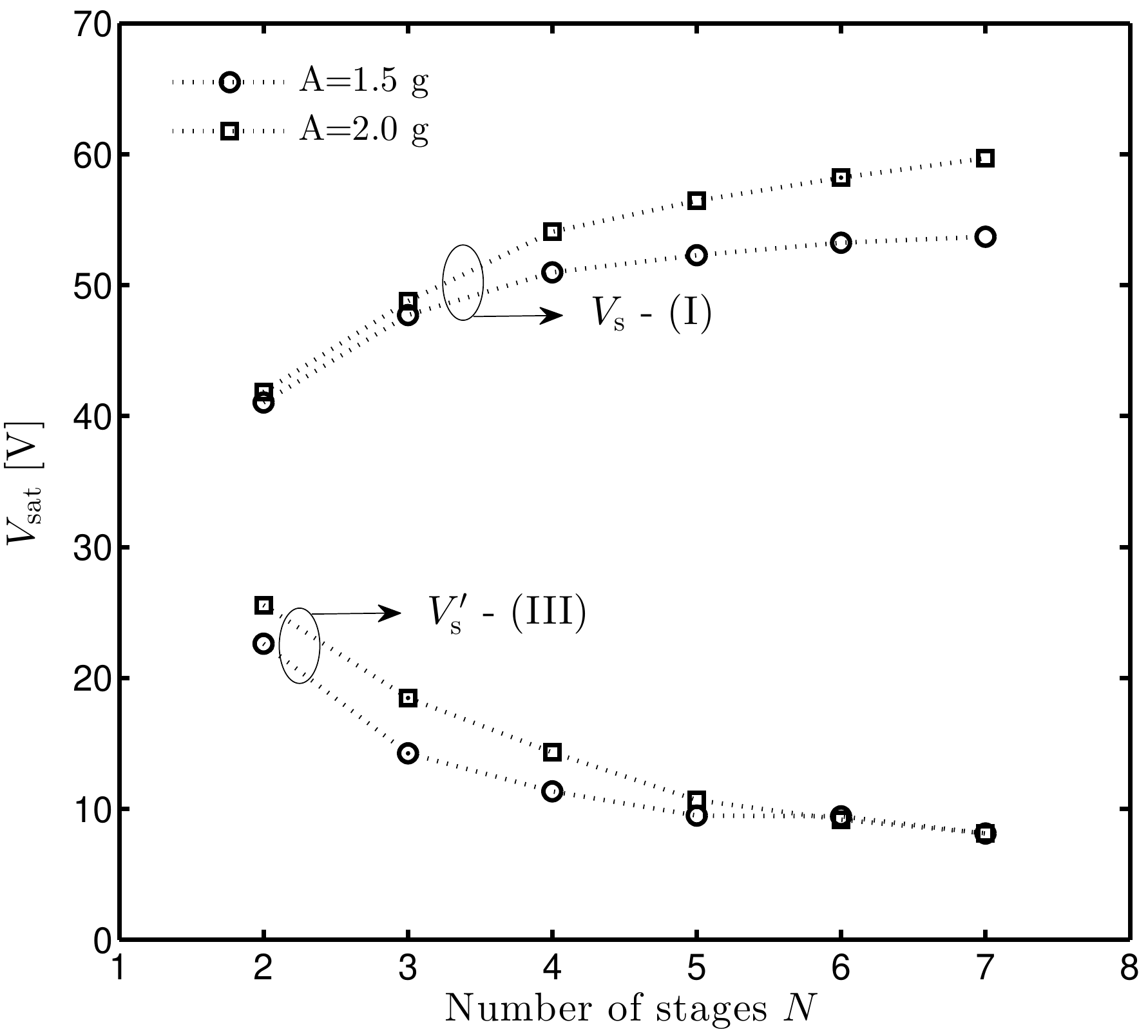}
	\caption{Saturation voltage $V_\mathrm{sat}$ versus number of stages $N$: comparison between configuration (I) and (III).}
	\label{fig_Vout_Comparison_1}
\end{figure}
\begin{figure}[!htbp]
	\centering
	\includegraphics[width=0.385\textwidth]{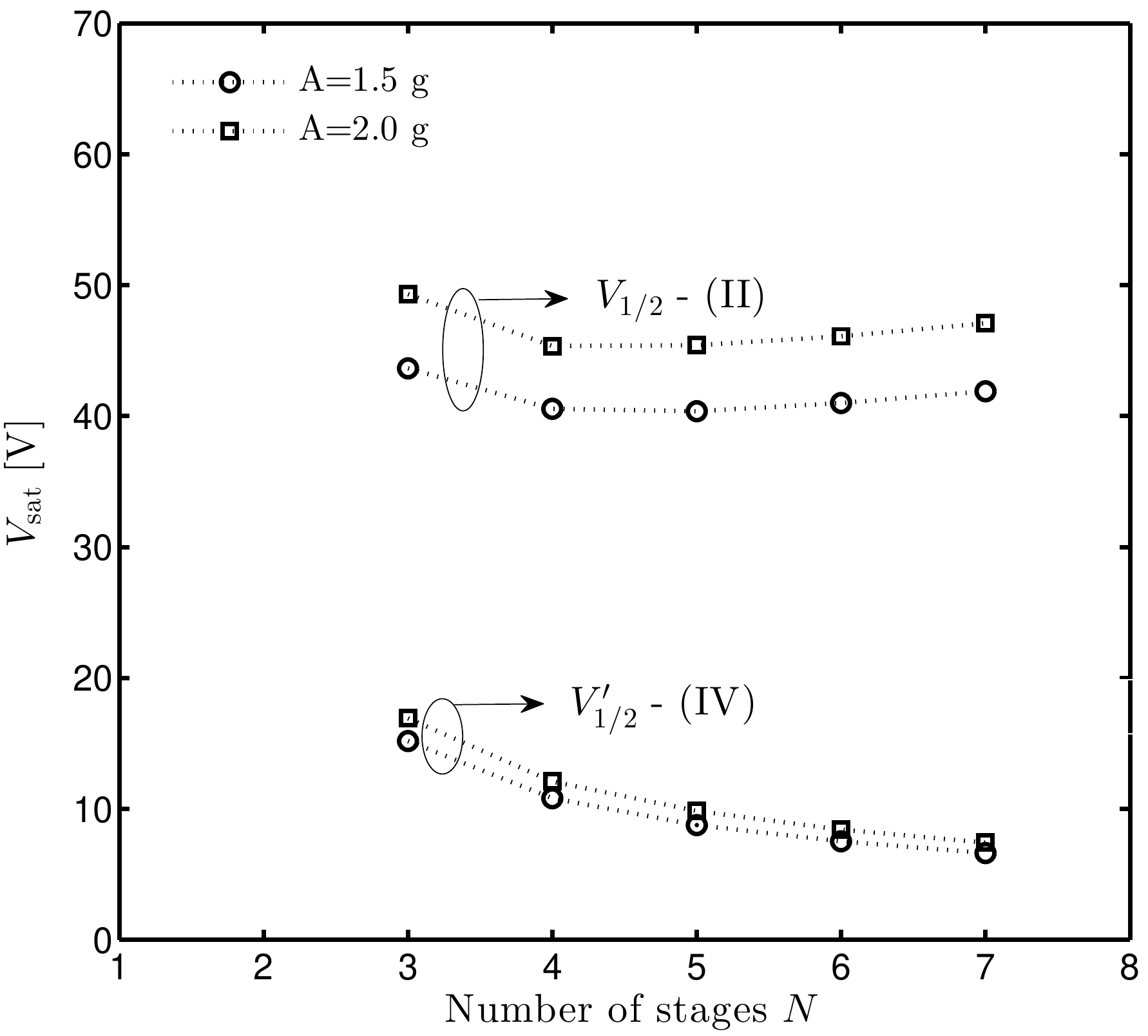}
	\caption{Saturation voltage $V_\mathrm{sat}$ versus number of stages $N$: comparison between configuration (II) and (IV).}
	\label{fig_Vout_Comparison_2}
\end{figure}
These above simulations are extended for various stage number $N$ with different input acceleration amplitudes. The saturation voltage results are presented in Figure \ref{fig_Vout_Comparison_1} and \ref{fig_Vout_Comparison_2}.
In both cases, the introduced topology shows the superiority in comparison with previous work by Lefeuvre \textit{et al.} Not only that, the remarkable gap of saturation voltage between the two topologies also raises with increase of $N$. The number of stage has a strong influence on the saturation voltage of the circuits. For configurations (III) and (IV), $V_\mathrm{sat}$ decreases as $N$ increases due to parasitic effect of the constituent capacitor of each stage and power losses in diodes. Though the configuration (I) can overcome this challenge, the voltage gain is less than linear as more stages are introduced. The behavior of the configuration (II) is a bit more complicated, however, the voltage changes are small in general.

The difference in performance between the configurations (I) and (III) can be explained by their physical mechanisms. For configurations (I), each stage is a modified voltage doubler, it also acts as an intermediate charge reservoir. An amount of energy stored in stage $j-\mathrm{th}$ is transferred into stage $(j+1)-\mathrm{th}$ in the next operation cycle, which keeps going until it reaches and then is stored in $C_\mathrm{s}$. Meanwhile, only a small amount of charge is transferred from $C_\mathrm{s}$ into $C_1$, making the voltage across $C_1$ to be built up very slowly. Considering configuration (III), in contrary, the harvested energy is pumped from $C_1$ to $C_\mathrm{s}$ through the intermediate capacitor $C_2$ alone. However, both $C_1$ and $C_2$ are charged by a built-in voltage multiplier circuit that may include up to $N$ cells. This process causes the voltages across the two variable capacitors to increase too fast, leading to rapid rise of the electrostatic force and decrease of the proof mass displacement as a consequence. Hence, the output voltage saturates at a lower level. That is also the reason why configuration (I) tend to have longer start-up time than configuration (III). Similar behaviors can be correspondingly inferred for the other configurations (II) and (IV).

\begin{table} [!thb]
	\centering
	\caption{Comparison of minimum required $\eta$ and $V_0$ between four configurations with $N=2$}%
	\begin{tabular}{l | l l | l l}
		\hline\hline
		\textbf{Configuration} & \textbf{(I)} & \textbf{(III)} & \textbf{(II)} & \textbf{(IV)} \\
		\hline
		$\eta_\mathrm{min}$ & 1.55 & 1.44 & 1.68 &  1.62 \\
		$(V_0)_\mathrm{min}$ & 6.0 & 4.5 & 10.0 &  10.0 \\
		\hline\hline
	\end{tabular}
	\label{Tab_V0_Eta_Min} 
\end{table}
The minimum required values of $\eta$ and $V_0$ for four configurations are compared in pairs in Table \ref{Tab_V0_Eta_Min}. Since the differences between them are not so significant, advantages of the proposed topologies are not overshadowed.

\begin{figure}[!htbp]
	\centering
	\includegraphics[width=0.3\textwidth]{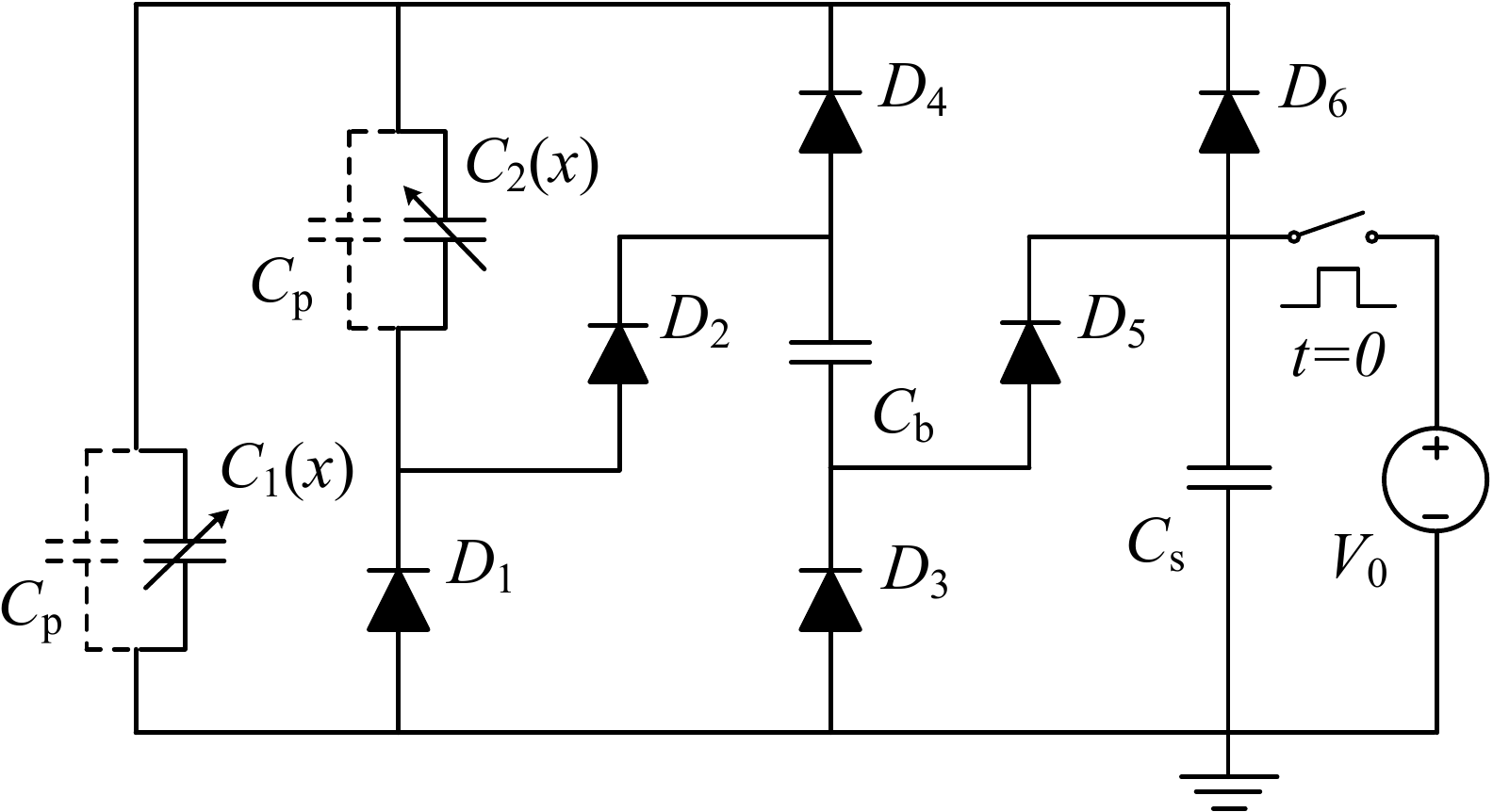}
	\caption{Series-parallel charge-pump circuit with $N=2$, by Karami \textit{et al.} \cite{Karami2017}.}
	\label{fig_Karami}
\end{figure}
\begin{figure}[!htbp]
	\centering
	\includegraphics[width=0.385\textwidth]{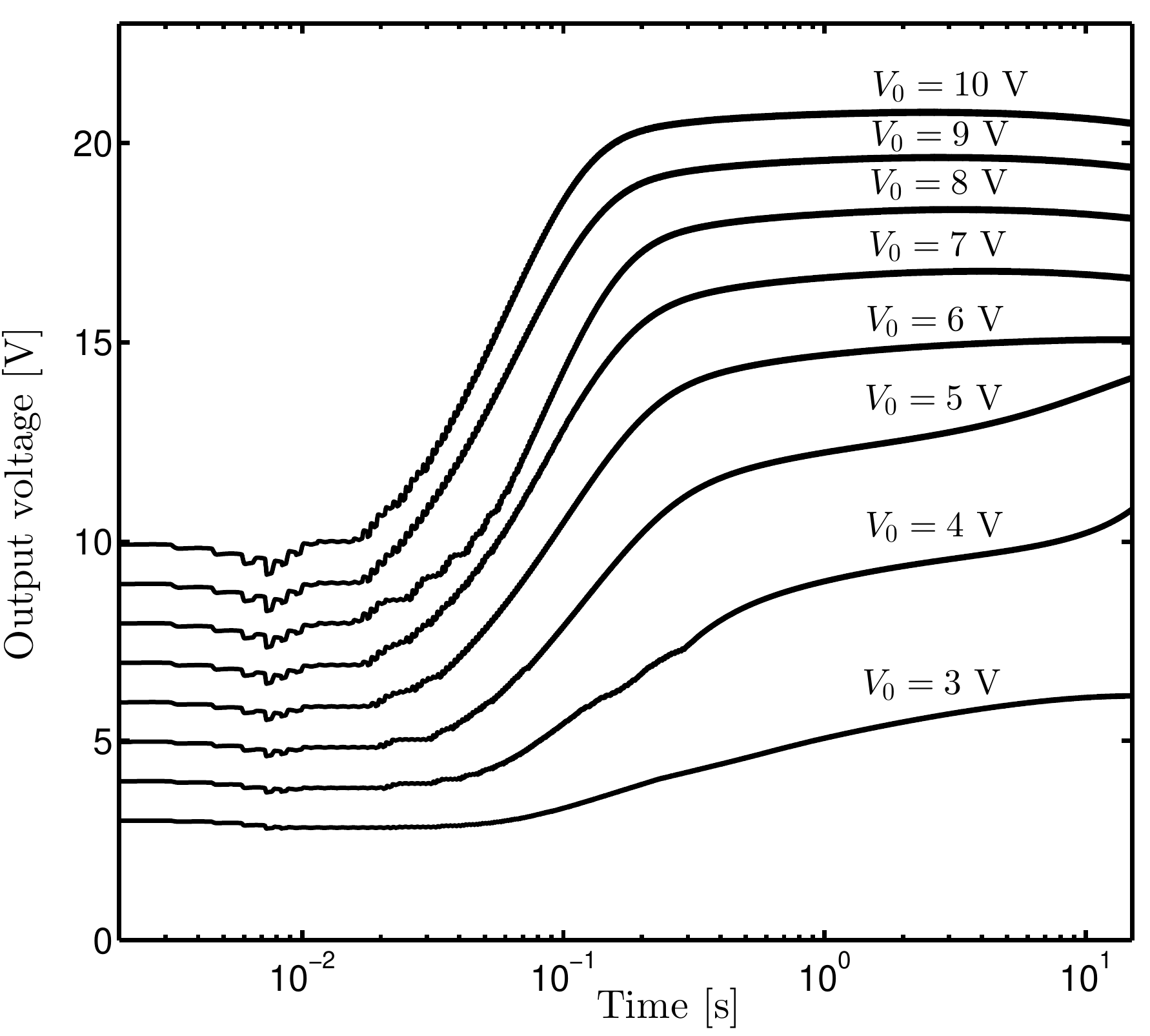}
	\caption{Evolution of the output voltage for different values of the initial bias $V_0$ ranging from 3 to 10 V.}
	\label{fig_Vout_Karami}
\end{figure}
The topology reported by Karami \textit{et al.} is shown in Figure \ref{fig_Karami} for $N=2$, adapted from \cite{Karami2017}. The same switch-capacitor configuration can be found in \cite{Dhulst2005}.
Simulation results show that the circuit cannot operate with $\eta=1.55$. Instead, the minimum required ratio of the capacitance variation is much higher than that, $\eta_\mathrm{min}=2.1$. Unlike other voltage multiplier circuits, the saturation voltage here strongly depends on the initial bias, but the voltage gain $V_\mathrm{sat}/V_0$ is very small, e.g., $V_\mathrm{sat} \approx 22$ V when $V_0=20$ V. Moreover, $V_\mathrm{sat}/V_0 > 1$ only if $V_0 > 6$ V.

The saturation voltage could be significantly improved when the device with high ratio of $\eta = 2.61$ is used. The corresponding maximum displacement is now $X_\mathrm{max}=11$ $\mu$m. The time evolution of the output voltage across the storage capacitor $C_\mathrm{s}$ is presented in Figure \ref{fig_Vout_Karami} with various values of the initial bias $V_0$. Since higher $V_0$ results in higher $V_\mathrm{sat}$, the investigation of the minimum pre-charge voltage is irrelevant in this case. Such a phenomenon has not been revealed in \cite{Karami2017} since only electrical domain was investigated. These results confirm the essential role of the electromechanical coupling on the circuit performance, which cannot be ignored in studying the system dynamics.

\section{Conclusion}
This paper presents a new configuration of a voltage multiplier for MEMS electrostatic energy harvesters.
The anti-phase gap-closing transducers were modeled and utilized in the study.
Without switches and inductive elements, the circuit proved to be suitable for ultra-low power implementation. Effect of the component imperfections such as parasitic capacitance of the transducer and diode non-idealities on the system performance are taken into account. 
The influence of biasing capacitances on start-up time and the dependence of the ripple voltage on the storage capacitor were explored.
The analyses shown that the saturation voltage depends on both the nonlinearities of the electrostatic force and the impact force at high acceleration amplitudes.
In comparing to previously reported circuits that are also based on the voltage doubler, the alternative topology presented in this paper can operate with a ratio of capacitance $\eta$ lower than 2. The minimum value of $\eta$ can be reduced by increasing the number of doubler stages $N$.
The introduced configuration has shown the superiority on the saturation voltage compared to earlier work published in the literature.

With the assessment that the electrostatic force of the in-plane overlap-varying harvester is proportional to the proof mass displacement (i.e., which is very different from the anti-phase gap-closing devices), performance of such a transducer structure when configured as N-stage voltage multiplier is interesting for further investigation.



%




\section*{Acknowledgment}
Financial support from the Research Council of Norway through Grant no. 229716/E20 is gratefully acknowledged.

\ifCLASSOPTIONcaptionsoff
  \newpage
\fi

\begin{IEEEbiography}{Binh Duc Truong}
	received the B.E. degree in Mechatronics from the Ho Chi Minh City University of Technology, Vietnam, in 2012
	and the M.Sc. degree in Micro- and Nano System Technology from the Buskerud and Vestfold University College, Norway, in 2015. He is currently pursuing the Ph.D. degree with Department of Microsystems, University College of Southeast Norway, focusing on MEMS electrostatic energy harvesters and wireless power transfer systems.
\end{IEEEbiography}


\begin{IEEEbiography}{Cuong Phu Le}
	Cuong Phu Le received the B.S. degree in telecommunications engineering from the University of Technology, Ho Chi Minh City, Vietnam, in 2005, the M.S. degree in microsystem technology from Vestfold University College in Horten, Norway in 2009 and the Ph.D. degree in microsystems technology from University of Oslo, Norway in 2013. He is currently a research scientist in Department of Microsystems, University College of Southeast Norway in Horten. His current research interests include microsystem design, modelling and MEMS vibration energy harvesting.    
\end{IEEEbiography}


\begin{IEEEbiography}{Einar Halvorsen}
	(M'03) received the Siv.Ing. degree in physical electronics from the Norwegian Institute of Technology (NTH), Trondheim, Norway, in 1991, and the Dr.Ing. degree in physics from the Norwegian University of Science and Technology (NTNU, formerly NTH), Trondheim, Norway, in 1996. He has worked both in academia and the microelectronics industry. Since 2004, he has been with University College of Southeast Norway in Horten, Norway, where he is a professor of micro- and nanotechnology. His current main research interest is in theory, design, and modelling of microelectromechanical devices. 
\end{IEEEbiography}


\vfill


\end{document}